\documentclass[lettersize,journal]{IEEEtran}

\usepackage{tikz}
\usepackage{amsmath}

\usepackage[linesnumbered, ruled]{algorithm2e}

\usepackage{colortbl}
\usepackage{eucal}
\usepackage{bm}
\usepackage{subfigure}
\usepackage{booktabs}
\usepackage{multirow}
\usepackage{enumitem}
\usepackage{url}
\usepackage{xspace}
\usepackage{amsfonts}
\usepackage{booktabs}
\usepackage{algorithmic}

\usepackage{booktabs} 
\usepackage{diagbox}    
\usepackage{makecell} 
\usepackage{float}
\usepackage{pifont}

\usepackage{graphicx} 
\usepackage{booktabs}
\usepackage{pifont}
\usepackage{multirow}

\newcommand{\mixedmark}{\ding{51}\textsuperscript{\kern-0.7em\raisebox{-0.6ex}{\tiny\ding{56}}}}

\newcommand{\rev}[1]{\textcolor{black}{#1}}

\newcommand{\revtwo}[1]{\textcolor{black}{#1}}

\newcolumntype{C}[1]{>{\centering\arraybackslash}p{#1}}
\newcolumntype{L}[1]{>{\raggedright\arraybackslash}p{#1}}
\newcolumntype{H}{>{\setbox0=\hbox\bgroup}c<{\egroup}}

\def\ourSystem{{FuseFi}\xspace}

\def\eg{\textit{e.g.},\xspace}
\def\etl{\textit{et al.}\xspace}

\begin{document}

\title{\rev{FuseFi}: Combining Irregularly Sampled CSI from Diverse Communication Packets and Frequency Bands for Wi-Fi Sensing}

\author{Gaofeng Dong, Kang Yang, and Mani Srivastava,~\IEEEmembership{Fellow,~IEEE}
\thanks{The authors are with the Department of Electrical and Computer Engineering, University of California, Los Angeles, CA, USA (email: {gfdong, kyang73, mbs}@ucla.edu).}
\thanks{Mani Srivastava holds concurrent appointments as a Professor of ECE and CS (joint) at the University of California, Los Angeles.}
\thanks{© 2026 IEEE.  Personal use of this material is permitted.  Permission from IEEE must be obtained for all other uses, in any current or future media, including reprinting/republishing this material for advertising or promotional purposes, creating new collective works, for resale or redistribution to servers or lists, or reuse of any copyrighted component of this work in other works.}
}

\maketitle


\begin{abstract}

Existing Wi-Fi sensing systems rely on injecting high-rate probing packets to extract channel state information~(CSI), leading to communication degradation and \rev{limited deployment flexibility}.
Although Integrated Sensing and Communication (ISAC) is a promising direction, existing solutions still rely on auxiliary packet injection because they exploit only uniform CSI from a single frame type, discarding approximately 70\% of naturally available packets.
We present \ourSystem, a novel Wi-Fi-based ISAC framework that directly exploits irregularly sampled CSI from diverse communication packets across multiple frequency bands, eliminating intrusive packet injection \rev{and introducing no sensing-specific communication overhead}.
\ourSystem integrates a CSI sanitization pipeline to harmonize heterogeneous packets and remove burst-induced redundancy, together with a time-aware attention model that learns directly from non-uniform CSI sequences without resampling.
We further introduce CommCSI-HAR, a new dataset with irregularly sampled CSI from real-world dual-band communication traffic.
\revtwo{Extensive evaluations on this dataset and, separately, on four public sensing tasks across three benchmark datasets show that \ourSystem achieves state-of-the-art accuracy with a compact model size, while fully preserving communication throughput.}

\end{abstract}

\begin{IEEEkeywords}
Channel state information (CSI), Internet of Things (IoT), Wi-Fi Sensing, Integrated Sensing and Communication (ISAC)
\end{IEEEkeywords}

\section{Introduction}\label{sec_introduction}

Recent years have witnessed extensive efforts on developing Wi-Fi sensing systems~\cite{wang2023automatic,xu2023self}. 
Leveraging Wi-Fi Channel State Information~(CSI), these systems enable ubiquitous sensing for a wide range of applications, including indoor positioning~\cite{niu2021understanding}, human activity recognition~\cite{deng2022gaitfi}, and identification~\cite{liu2024unifi}.
However, existing Wi-Fi sensing systems often rely on specific radio configurations and specially crafted probing packets to achieve sufficiently high CSI sampling rates, typically 100–1000 Hz~\cite{cominelli2023exposing,ji2023construct,zhang2019commercial,shi2020towards,zhang2022handgest}. 
While this ensures adequate fixed-rate CSI data for existing sensing models designed around fixed-length inputs, it comes at a cost: injected packets interfere with normal communication, reducing throughput by over 40\%~\cite{yang2022wiimg}, and may collide with ongoing transmissions, further degrading sensing performance~\cite{guo2022twcc,yang2022wiimg}. 

To address this challenge, SenCom~\cite{he2023sencom} proposed a practical Wi-Fi-based Integrated Sensing and Communication~(ISAC)~\cite{chen2022wi,restuccia2021ieee,liu2022integrated} system, which introduces a calibration process to extract CSI from data packets in communication traffic.
However, SenCom only utilizes packets of a single type in a single frequency band, thereby discarding over 70\% of available non-data packets based on our empirical analysis in Section~\ref{subsec:empirical_analysis}.
Furthermore, it lacks a model capable of directly processing irregularly sampled CSI sequences of varying lengths. Consequently, it relies on an incentive mechanism to sustain a fixed-rate sampling for downstream learning models. As a result, it still requires injecting redundant dummy packets when the data packet rate is below 100~Hz, preventing the complete elimination of sensing-induced overhead.

\begin{figure}[t]
    \centering
    \includegraphics[width=\linewidth]{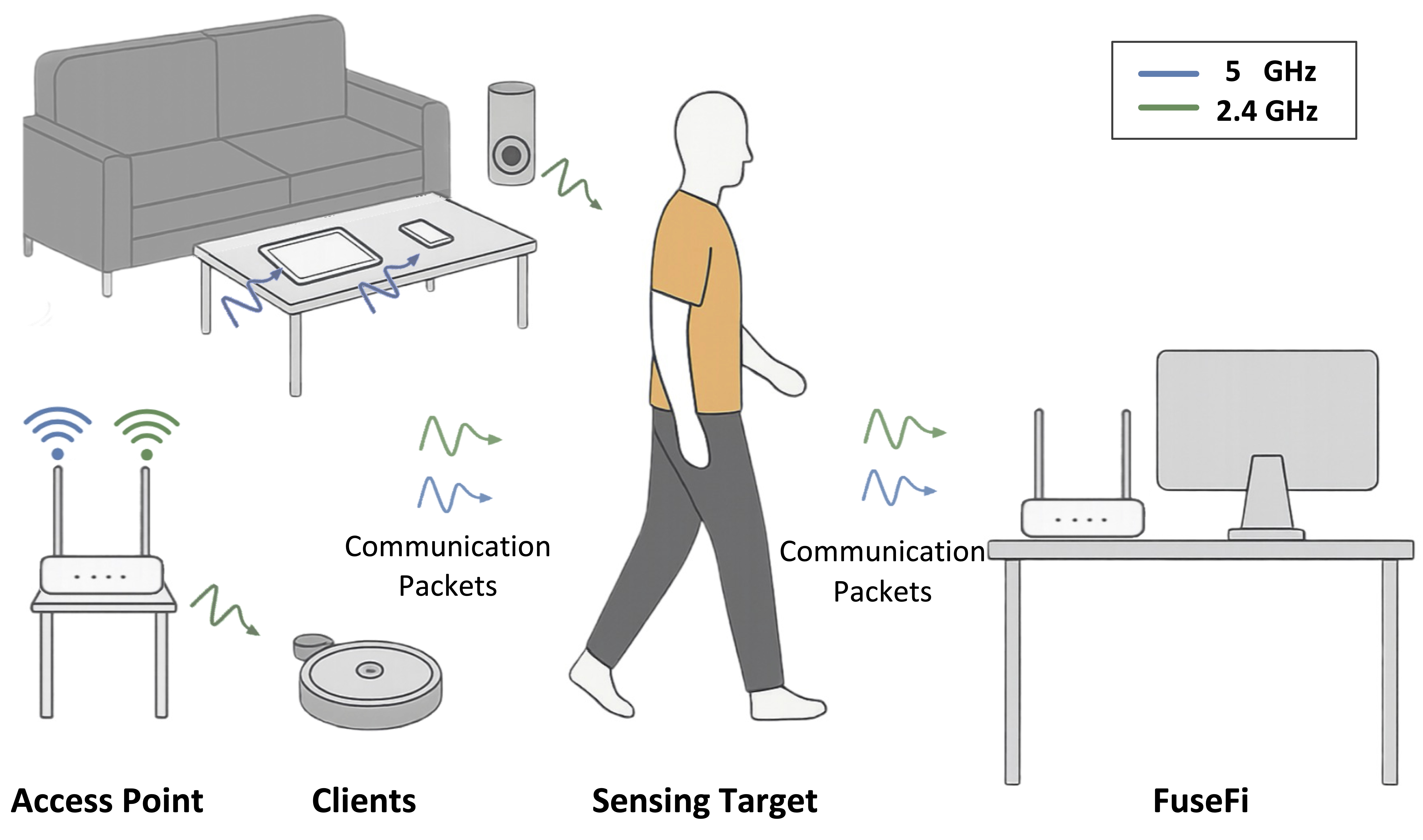}
    \caption{Application scenario of \ourSystem. It utilizes irregularly sampled CSI from diverse communication packets across multiple frequency bands for sensing.}
    \label{fig:scenario}
\end{figure}

This paper presents \ourSystem, a novel ISAC framework that directly exploits irregularly sampled CSI from communication traffic without intrusive packet injection, \rev{enabling a practical and efficient Wi-Fi sensing system that could be integrated into existing communication systems when firmware- or driver-level CSI exposure is available}, as shown in Figure~\ref{fig:scenario}.
Intuitively, this is feasible because modern Wi-Fi networks naturally generate abundant and diverse communication packets, including data, management, and control frames, across multiple frequency bands, providing a rich source of CSI variations for sensing.
For example, commodity devices support dual-band operation on 2.4 and 5\,GHz, and Wi-Fi~7 Multi-Link Operation (MLO) \rev{ further reflects the industry trend toward simultaneous multi-band connectivity}.
Existing works do not fully utilize these naturally available packets~\cite{zhang2023wifi}.

\begin{figure*}[!ht]
    \centering
    \includegraphics[width=0.9\linewidth]{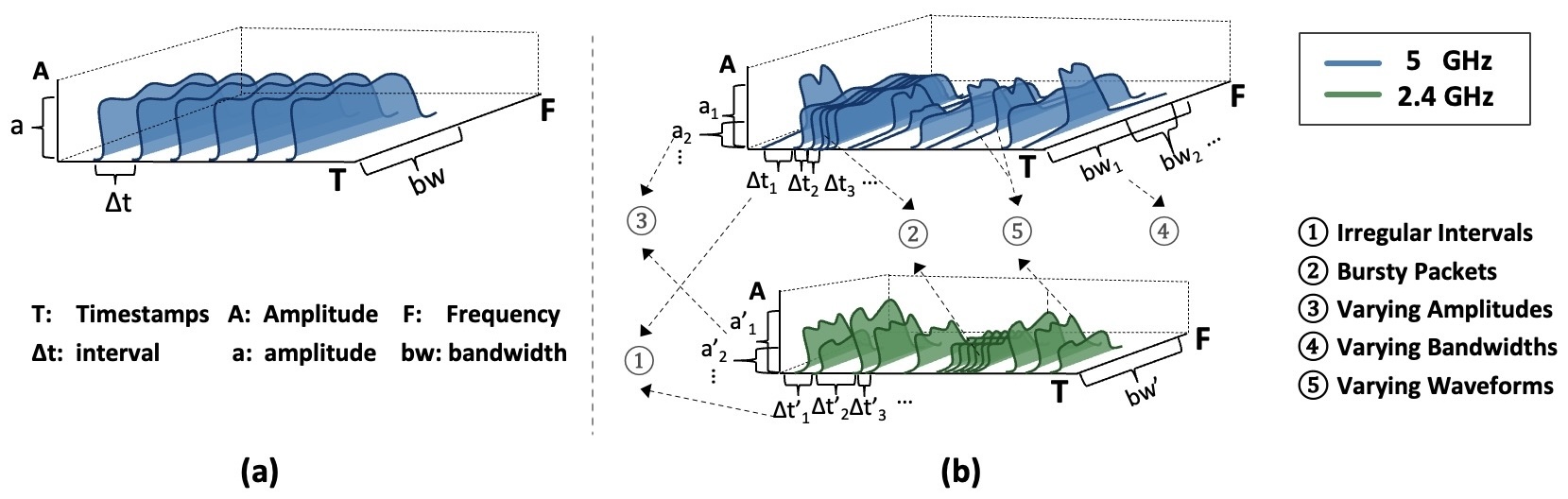}
    \caption{CSI comparison: traditional Wi-Fi sensing frameworks vs. \ourSystem. 
    \textbf{(a)} Traditional frameworks inject controlled packets for uniform CSI with constant intervals, bandwidths, and amplitudes. 
    \textbf{(b)} Our \ourSystem framework leverages CSI from multi-band communication packets with irregular intervals, bursty timestamps, dynamic amplitudes, variable bandwidths, and different waveforms.}
    \label{fig:challenges}
\end{figure*}

Although promising, \ourSystem faces new challenges, as shown in Figure~\ref{fig:challenges}: 
{\textit{First,}} communication packets are inherently irregular and bursty, with rates varying from 10 Hz to several kHz.\footnote{The default Beacon Frame interval is 100 Time Units~(TUs), where one TU equals $1024~\mu s$~\cite{wikipedia_beacon_frame}. This corresponds to approximately 10 Hz.} Such variability produces CSI sequences of different lengths, which existing Wi-Fi sensing frameworks cannot handle because they mainly rely on deep neural networks (DNNs) that require fixed-length, regularly sampled inputs~\cite{yang2022wiimg,cominelli2023exposing,yang2023sensefi}. Moreover, bursty CSI often contains redundant information about the channel within short coherence periods, and thus must be identified and filtered out.  
{\textit{Second,}} unlike injected or probing packets with controlled configurations, communication-driven adaptations in power introduce CSI variations unrelated to the sensing target, which must be properly handled before being used by deep learning models. 
{\textit{Third,}} communication packets include diverse frame types whose bandwidths and waveforms differ, complicating CSI processing. A simple workaround is to restrict sensing to a single frame type~\cite{he2023sencom}, but this approach disregards many other frames that also carry valuable sensing information.  
{\textit{Fourth,}} existing Wi-Fi sensing datasets~\cite{yousefi2017survey, zhang2021widar3, meneghello2022sharp, yang2023sensefi, cominelli2023exposing, wang2024xrf55, yang2023mm, yuanoctonet, zhu2025csi} collect CSI from injected fixed-rate packets with uniform frame types and bandwidths, as illustrated in Fig.~\ref{fig:challenges}(a). 
No public dataset captures CSI from normal \emph{communication} traffic with the heterogeneous waveforms, variable bandwidths, and irregular timestamps that characterize real ISAC scenarios.

To address these challenges, \ourSystem introduces the following advances. 
Instead of relying on a single frame type, we incorporate CSI from all data, management, and control frames, sanitizing them through a dedicated signal processing pipeline. 
We further filter bursty CSI, ensuring efficient and non-redundant sensing. 
For processed irregularly sampled CSI with varying input lengths and bandwidths, we employ a time-aware attention model that embeds continuous timestamps and variable channels to generate fixed-length representations.
Moreover, leveraging the multi-band capabilities of commodity Wi-Fi devices, such as the 2.4 and 5~GHz bands, \ourSystem naturally fuses multi-band CSI for robust sensing, even when the measurements are irregularly sampled and derived from diverse packet types.
To address the dataset gap, we further introduce CommCSI-HAR, a human activity recognition (HAR) dataset collected using COTS devices that, to the best of our knowledge, is the first to contain irregularly sampled CSI derived from diverse packet types in real dual-band communication traffic.

Extensive evaluations on this dataset show that \ourSystem{} outperforms existing baselines and achieves the highest accuracy among communication-CSI settings, while \rev{performing competitively with} the fixed-rate reference setting, whose CSI is collected concurrently from injected packets at a fixed rate.
\revtwo{To further evaluate the architectural generality of \ourSystem across different sensing tasks and dataset settings,}
we also test it on four public sensing tasks drawn from three benchmark datasets: Gesture Recognition~\cite{zhao2024finding}, Action Recognition and Fall Detection~\cite{zhao2024knn}, and People Counting~\cite{zhao2024mining}, where CSI is still obtained from injected fixed-rate packets of a single type and bandwidth, but with \textit{minor packet loss}.
Across all datasets, \ourSystem achieves state-of-the-art performance while using only around 3\% parameters of the baseline, \rev{demonstrating its effectiveness and efficiency across diverse sensing tasks and dataset settings}.
\rev{We believe \ourSystem provides an initial step toward practical Wi-Fi-based ISAC systems and can motivate future research on fully integrated implementations and robust real-world deployment.}

Our key contributions are summarized as follows: 

\begin{itemize}[label=\textbullet, leftmargin=1em, topsep=0em]
\item We introduce the task of Wi-Fi sensing using irregularly sampled CSI from diverse communication packets and bands, and present \ourSystem. \rev{To our knowledge, \ourSystem{} is the first Wi-Fi-CSI ISAC framework to systematically exploit irregularly sampled, naturally occurring multi-frame-type communication CSI across packet types and frequency bands without active packet injection.}

\item We propose a sanitization pipeline that maximizes the utility of CSI from communication traffic, coupled with a time-aware attention architecture that learns directly from irregularly sampled CSI, enabling end-to-end sensing tasks and scaling across different bandwidths and waveforms.

\item We further contribute a new HAR dataset with irregularly sampled CSI from diverse communication packets across dual frequency bands.
\footnote{Project webpage: \url{https://nesl.github.io/FuseFi/}.}

\item Comprehensive experiments across five sensing tasks, including HAR, gesture recognition, action detection, fall detection, and people counting, show that \ourSystem consistently achieves state-of-the-art performance with efficient model size and computation cost compared to baselines. 

\end{itemize}

\section{Background and Related Work}\label{sec_back}

\subsection{Wi-Fi Channel State Information }
\label{subsec:csi}

Channel State Information (CSI) is a physical layer metric that characterizes the propagation of wireless signals from a transmitter (Tx) to a receiver (Rx)~\cite{gast2013802, ma2019wifi, niu2021understanding}, providing a rich fingerprint of the environment. 
Modern Wi-Fi uses Orthogonal Frequency-Division Multiplexing (OFDM), which partitions a wide, frequency-selective channel into many narrow subcarriers that are approximately flat-fading. CSI is estimated per packet from known training symbols embedded in the packet preamble, i.e., the Long Training Field (LTF). 
At the Rx, the baseband model on subcarrier $k$ is:
\begin{equation}
\mathbf{y}_k = \mathbf{H}_k \mathbf{x}_k + \mathbf{n}_k
\end{equation}
where $\mathbf{x}_k\in\mathbb{C}^{N_t}$ is the vector of known training symbols transmitted across $N_t$ Tx antennas, $\mathbf{y}_k\in\mathbb{C}^{N_r}$ is the corresponding received vector across $N_r$ Rx antennas, $\mathbf{H}_k\in\mathbb{C}^{N_r\times N_t}$ is the per-subcarrier CSI matrix, and $\mathbf{n}_k$ models additive noise. 
For a single packet, aggregating the estimates across all $K$ subcarriers forms a three-dimensional complex CSI tensor, $\mathbf{H} \in \mathbb{C}^{N_r \times N_t \times K}$, whose entry $H_{i,j,k}$ is the CFR between Tx antenna $j$ and Rx antenna $i$ on subcarrier $k$. 
Over time, a stream of CSI yields a fourth-order structure, enabling temporal analysis of channel dynamics.
In this work, we focus on CSI amplitude, since phase is generally more sensitive to noise and synchronization imperfections~\cite{xie2015precise,hu2024you}. The implications of this design choice, along with the potential for incorporating phase information, are further discussed in Section~\ref{sec:discussion}.

\subsection{Related Work}\label{sec_relatedwork}

\textbf{CSI-based Sensing:} Most existing Wi-Fi CSI-based sensing systems rely on extracting high fixed-rate CSI by injecting dummy packets at high rates (\eg 100-1000 Hz)~\cite{cominelli2023exposing, li2016wifinger, zhang2019commercial, shi2020towards, guo2022twcc}, which causes network congestion and disrupts normal communication. 
Some studies~\cite{zhang2019widetect, yu2025lowdetrack, zhang2019smars} use lower fixed-rate CSI (\eg 30 Hz), but this reduces the performance.
To cope with low rates, interpolation techniques~\cite{palipana2018falldefi, gao2021towards, zheng2016smokey} are proposed, but they fail to fully leverage data correlation and often amplify noise, as noted in~\cite{zheng2024pushing}. 
Then Zheng \etl~\cite{zheng2024pushing} recover high-rate CSI from low-rate CSI (\eg 25/50/100 Hz) using GANs, but this approach still focuses on fixed-rate intervals, limiting its applicability in real-world settings where CSI rates fluctuate. 
Additionally, it discards abundant useful CSI when the rates fall between these intervals and does not account for the irregular nature of real-world communications. 
More recently, Abedi \etl~\cite{abedi2023wifi} inject fake packets to trigger ACKs and achieve higher CSI rates (70~Hz), but they only use ACK frames for sensing, ignoring other valuable frame types.
In this paper, \ourSystem directly leverages the diverse communication packets already present in Wi-Fi traffic for sensing, eliminating the need for packet injection, despite their irregular intervals.

While the majority of Wi-Fi sensing systems rely on fixed-rate CSI inputs~\cite{zheng2024pushing, shi2020towards, guo2022twcc}, CSI-BERT2~\cite{zhao2024mining} addresses the issue of variable sequence lengths caused by packet loss, which can reach 14.5\% when injecting packets at 100~Hz. To manage these incomplete sequences, it employs a placeholder [PAD] to fill gaps to reconstruct padded inputs for its BERT-based backbone. 
Although CSI-BERT2 is not explicitly designed for ISAC systems, 
we treat it as an underlying model capable of processing CSI sequences with varying effective lengths.

\noindent \textbf{Multiband CSI Combination:} A parallel line of work combines CSI from multiple narrow frequency bands to synthesize a wider effective bandwidth for improved range resolution. ToneTrack~\cite{xiong2015tonetrack} and $M^3$~\cite{chen2019m} concatenate CSI across adjacent Wi-Fi channels for sub-meter localization, relying on phase alignment within short time windows and high packet rates (200\,pkt/s). HiSAC~\cite{pegoraro2024hisac} extends this to non-contiguous subbands for ranging of mainly stationary objects, though moving targets require inter-packet intervals of 5\,ms (200\,pkt/s). $\mu$Ceiver-Fi~\cite{li2025muceiver} exploits multi-link receivers for GHz-level bandwidth synthesis and fine-grained pose estimation, also at 200\,Hz.
A common thread across these works is that they all assume cooperative and/or high-rate transmissions, either controlled packet injection or densely sampled traffic within short coherence windows, so that CSI from different bands can be phase-aligned and stitched in the frequency domain. 
FuseFi addresses a fundamentally different challenge: extracting temporal sensing cues from ambient communication traffic whose packet types, bandwidths, amplitudes, and inter-arrival times are all uncontrolled and highly irregular. 
Rather than synthesizing a coherent wideband channel estimate, FuseFi treats multi-band and multi-type CSI as complementary but independently informative observations, fusing them through a learned time-aware attention mechanism that operates directly on irregular timestamps.

\noindent \textbf{ISAC:} 
Existing ISAC research primarily focuses on designing novel PHY layers that are suitable for both communication and sensing~\cite{liu2022integrated, zhang2024integrated, li2025ris, lu2024integrated}. 
Additionally, the new amendment to the Wi-Fi standard, 802.11bf, introduces a novel WLAN sensing procedure~\cite{du2024overview, restuccia2021ieee, chen2022wi}. 
However, these approaches require modifications to existing systems, making them incompatible with billions of existing Wi-Fi devices. 
In contrast, \ourSystem aims to achieve ISAC using COTS devices, ensuring compatibility with current Wi-Fi infrastructure.

SenCom~\cite{he2023sencom} is an important prior practical ISAC system that extracts CSI from communication data packets for sensing. When the data packet rate drops below the targeted 100~Hz, it injects incentive probing packets to obtain more CSI, and uses fitting and resampling to obtain fixed-length CSI sequences. SenCom targets a different but related design point: its well-engineered target-rate calibration and fitting/resampling pipeline extracts sensing-grade CSI from communication packets under its target-rate framework, whereas \ourSystem targets learning directly from irregular, heterogeneous, multi-band communication CSI without packet injection.
In particular, SenCom focuses on CSI from a single frame type and is not able to directly process irregularly sampled CSI. \ourSystem utilizes a sanitization pipeline and a customized DNN architecture to process all available communication packets without active injections, providing a different solution to the injection-free Wi-Fi sensing problem.

\noindent \textbf{Comparison with \ourSystem:}
\rev{Table~\ref{table:compare_related_work} summarizes the capability and design-scope differences between \ourSystem{} and two representative systems: SenCom~\cite{he2023sencom}, an important practical ISAC system with a target-rate calibration and fitting/resampling pipeline, and CSI-BERT2~\cite{zhao2024mining}, a robust sensing model for varying-length CSI inputs. This comparison focuses on input assumptions and supported capabilities, since the systems target different sensing settings and were evaluated under different experimental setups.}

In terms of the system \textit{Setting}, \ourSystem stands out as an injection-free ISAC solution. While SenCom is designed for ISAC, it is marked with a hybrid checkmark for ``Injection-free'' because it still requires injecting redundant packets when the data packet rate falls below 100~Hz, introducing potential interference not only to the monitored communication link but also to other links sharing the same channel.

Regarding \textit{Data} diversity, \ourSystem is a framework that combines heterogeneous CSI sources, by fusing multi-type (data, management, control), multi-bandwidth, and multi-band CSI. This capability allows \ourSystem to exploit the rich variance in communication traffic, whereas SenCom and CSI-BERT2 are limited to a specific frame type and bandwidth.

Regarding \textit{Model} capabilities, we assign CSI-BERT2 a partial checkmark for adaptability. While it accommodates varying sequence lengths via padding, it introduces additional computation and redundancy, and may not handle the irregular, bursty timestamps and heterogeneous CSI inherent in real-world traffic. SenCom lacks a model capable of directly processing varying-length inputs; instead, it relies on fitting and resampling techniques to reconstruct fixed-rate sequences for standard sensing models. In contrast, \ourSystem introduces a time-aware attention architecture that learns directly from irregular timestamps and diverse CSI sequences.

Finally, regarding \textit{Dataset} contributions, to facilitate future research in this domain, we present a new dataset (CommCSI-HAR) collected from real \textit{communication} traffic, addressing the lack of public datasets for this ISAC scenario.

\begin{table}[t]
\small
\centering
\caption{\rev{Capability and design-scope comparison with related Wi-Fi sensing systems.}}
\begin{tabular}{llccc}
\toprule
 &  
 & \makecell{SenCom\\\cite{he2023sencom}} 
 & \makecell{CSI-BERT2\\\cite{zhao2024mining}} 
 & \textbf{\ourSystem} \\
\midrule

\multirow{2}{*}{Setting} 
&ISAC  & \ding{51} & \ding{55} & \ding{51} \\
&Injection-free & {\ding{51}\textsuperscript{\kern-0.7em\raisebox{-0.6ex}{\tiny\ding{56}}}} & \ding{55} & \ding{51} \\
\midrule

\multirow{3}{*}{Data} 
&Multi-type & \ding{55} & \ding{55} & \ding{51} \\
&Multi-BW & \ding{55} & \ding{55} & \ding{51} \\
&Multi-band & \ding{55} & \ding{55} & \ding{51} \\
\midrule

\multirow{1}{*}{Model} 
&Adaptability & \ding{55} & {\ding{51}\textsuperscript{\kern-0.7em\raisebox{-0.6ex}{\tiny\ding{56}}}} & \ding{51} \\
\midrule

\multirow{1}{*}{Dataset} 
&Comm-CSI  & \ding{55} & \ding{55} & \ding{51} \\

\bottomrule

\end{tabular}
\label{table:compare_related_work}
\end{table}

\section{Motivation}

\subsection{Empirical Analysis}
\label{subsec:empirical_analysis}

To assess the feasibility of \ourSystem, we conducted an empirical study on the composition of real-world Wi-Fi traffic. 
In typical smart homes and offices, many connected devices, such as smartphones, tablets, laptops, TVs, voice assistants, cameras, and IoT sensors, generate Wi-Fi activity across multiple bands~\cite{huang2020iot, mazhar2020characterizing, mainuddin2022iot}.
We established a testbed in a controlled environment designed to mimic a typical household~\cite{parks_smart_home}, comprising one router and some clients: three smartphones, three tablets, two laptops, and two security cameras (operating only on 2.4~GHz).\footnote{All equipment consisted of controlled devices operated in an isolated network in this traffic analysis study. Data collection was strictly limited to packet timestamps and frame types, so no personally identifiable information was stored.} We captured the packets across various common activities (e.g., video streaming, web browsing, gaming)~\cite{he2023sencom} and frequency bands.

\begin{table}[t]
\centering
\caption{The percentages of non-data packets for different scenarios.}
\begin{tabular}{llc}
    \toprule
    \textbf{Band} & \textbf{Scenario} & \textbf{Percentage} \\ 
    \midrule
    
    \multirow{5}{*}{5 GHz} 
        & All devices: idle & 85.41\%  \\
        & 1 device: video & 70.91\%  \\
        & 1 device: webpage & 69.61\%  \\
        & 1 device: game & 70.79\%  \\
        & 3: webpage, video, social media  & 71.34\%  \\
    \midrule

    \multirow{3}{*}{2.4 GHz} 
        & All devices: idle & 94.46\%  \\
        & 1 camera: streaming video & 83.95\%  \\
        & 2 cameras: streaming video & 79.48\%  \\

    \bottomrule
\end{tabular}
\label{table:data_vs_nondata_percentage}
\end{table}

Table~\ref{table:data_vs_nondata_percentage} summarizes the proportion of non-data packets versus data packets in these scenarios.
The results reveal a substantial, often overlooked reservoir of sensing data. On both bands, non-data traffic consistently constitutes approximately 70\% of the total packet volume, even during video streaming. This is due to the necessity of control and management signaling: the router frequently transmits Request-to-Send (RTS) frames to coordinate downlink Quality of Service (QoS) data, ACKs to confirm uplink client traffic, and management packets. Although data frames occupy the majority of \textit{airtime} due to their payload size, we emphasize \textit{packet percentage} here because each packet, regardless of duration, yields a valid CSI estimate for sensing.
These findings indicate that existing ISAC frameworks focusing exclusively on (QoS) data frames discard the vast majority of available channel information, significantly limiting their temporal resolution and robustness.

However, leveraging this abundant communication traffic is non-trivial. These packets are fundamentally different from the uniform, controlled probes used in traditional sensing systems. 
As we detail in the following sections, effectively utilizing this diverse stream requires not only overcoming significant signal processing hurdles inherent to natural communication traffic but also developing novel deep learning architectures capable of learning directly from irregular and heterogeneous inputs.

\subsection{Heterogeneous CSI and Irregular Timestamps}
\label{subsec:comm_csi}

While the CSI formation process in Section~\ref{subsec:csi} applies to any single packet, exploiting CSI from normal communication traffic introduces challenges absent in injected customized packets at a fixed rate. Real Wi‑Fi links adapt their PHY on a per‑packet basis and generate traffic only when needed, yielding dynamic amplitudes, heterogeneous waveforms, variable bandwidths, and irregular timestamps, as shown in Figure~\ref{fig:challenges}~(b).

\subsubsection{Heterogeneous CSI}
\textbf{Diverse PHY formats.}
Wi‑Fi carries \emph{data}, \emph{management}, and \emph{control} frames across multiple PHY layer generations (e.g., 802.11n/ac/ax/be), employing different preamble structures and long‑training fields (e.g., L-LTF, HT‑LTF, VHT‑LTF). As a result, the raw CSI extracted from different packets may exhibit waveform‑dependent distortions that are artifacts of the preamble rather than the physical environment. In addition, bandwidth allocations are dynamic: links switch among \(20/40/80/etc\)~MHz.
Consequently, the CSI tensor shape changes from packet to packet, which we denote as
\begin{equation}
\tilde{\mathbf{H}}^{(p)} \in \mathbb{C}^{N_r \times N_t \times K^{(p)}}
\end{equation}
where \(p\) indexes packets,
and \(K^{(p)}\) depends on the instantaneous PHY configuration. 
More generally, the measured CSI can be viewed as a shaped version of the underlying channel,
\begin{equation}
\tilde{\mathbf{H}}^{(p)} \;=\; \mathbf{S}^{(p)} \odot \mathbf{H}^{(p)} \;+\; \mathbf{E}^{(p)}
\end{equation}
where \(\mathbf{S}^{(p)}\) captures format‑ and bandwidth‑ dependent preamble effects, \(\odot\) is the element‑wise product, and \(\mathbf{E}^{(p)}\) aggregates residual estimation errors and impairments.

\noindent
\textbf{{Power adaptation.}}
CSI's amplitude is also influenced by mechanisms unrelated to the sensing target. Transmit power control and the receiver’s automatic gain control (AGC) adjust power on a per‑packet basis to maintain link quality, which introduce packet‑to‑packet amplitude scaling:
\begin{equation}
\tilde{\mathbf{H}}^{(p)} \;=\; \alpha^{(p)} \big(\mathbf{S}^{(p)} \odot \mathbf{H}^{(p)} \big) \;+\; \mathbf{E}^{(p)}
\end{equation}
where \(\alpha^{(p)}\) denotes an unknown or varying gain factor governed by AGC and transmit-power adaptation. This gain variation can dominate the subtle motion-induced changes of interest. Since most prior Wi-Fi sensing studies extract relatively uniform CSI from a single type of injected packet under highly controlled settings, the impact of such packet-dependent amplitude variation on practical Wi-Fi sensing remains underexplored.

\subsubsection{Irregular and bursty timestamps}
\label{subsubsec:irregular_timestamps}
Since communication packets are generated on-demand and scheduled by CSMA/CA, their arrival times \(t_p\) are fundamentally irregular. The inter-packet interval, denoted by \(\Delta t_p = t_p - t_{p-1}\), is highly variable, characterized by long gaps during idle periods and dense bursts under heavy traffic. This temporal variability is particularly pronounced for data frames. In contrast, management and control frames can provide a more regular signal, such as periodic Beacon Frames, which are typically broadcast at 10~Hz. This overall temporal irregularity violates the uniform-grid assumption in traditional Wi-Fi sensing models and can bias temporal features if the data is simply interpolated.

\subsection{Scope of \ourSystem}
\label{subsec:scope}
As established above, CSI extracted from communication packets violates the assumptions underlying most existing Wi-Fi sensing pipelines: timestamps are not uniformly sampled, inputs are no longer fixed-dimensional, and amplitudes are affected not only by motion. Basic interpolation, resampling, or padding can blur motion cues or introduce additional noise, causing traditional DNNs to underperform or fail~\cite{zheng2024pushing,zhao2024mining}.

At the same time, ambient communication traffic represents a vast and largely unexploited sensing resource. Our goal is therefore to design a unified framework that combines irregularly sampled CSI from diverse communication packets and frequency bands for practical ISAC systems, enabling sensing from normal communication traffic without active packet injection. 
Crucially, FuseFi operates on per-packet CSI or equivalent channel estimates \rev{that Wi-Fi receivers already compute for normal decoding, so no additional radio resources, airtime, or sensing-specific packets are required on the monitored channel}. 
Deployment therefore reduces to exposing these channel estimates through firmware or driver support, as demonstrated by prior work such as Nexmon~\cite{schulz2018nexmon}, enabling integration into existing Wi-Fi devices.
The broader scope and future extensions of \ourSystem are discussed in Section~\ref{sec:discussion}.

\section{System Design}
\label{Sec:FuseFi-FuseFi}

In this section, we first present an overview of \ourSystem and then introduce the design details of each component.

\begin{figure*}[t]
    \centering
    \includegraphics[width=1.0\linewidth]{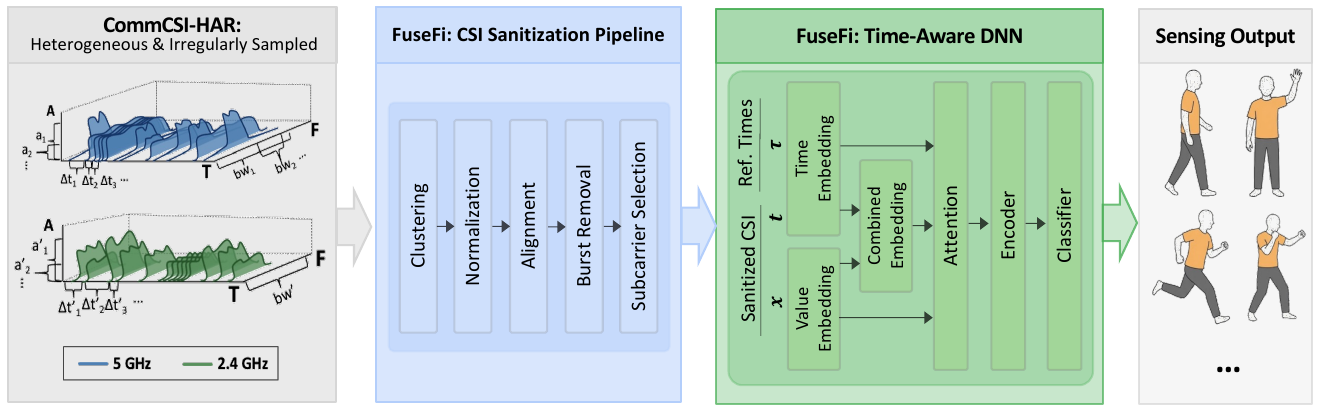}
    \caption{Architecture of \ourSystem. Taking the heterogeneous and irregularly sampled CSI (CommCSI-HAR dataset) as input, \ourSystem applies a CSI Sanitization pipeline to align waveforms, remove bursts, and select subcarriers. The processed data is then fed into a time-aware attention network to generate an output for the sensing tasks.}
\label{fig:system_overview}
\end{figure*}

\subsection{System Overview}
\label{subsec:overview}

\noindent
{FuseFi} sanitizes irregularly sampled and heterogeneous CSI from normal Wi-Fi communication into a unified representation suitable for learning-based sensing. As illustrated in Figure~\ref{fig:system_overview}, FuseFi consists of two stages: \emph{(1) CSI Sanitization Pipeline} and \emph{(2) Time-Aware DNN}. Given an input stream
\begin{equation}
\mathcal{S} = \{(\tilde{\mathbf{H}}^{(p)}, t_p, \text{meta}_p)\}_{p=1}^{T},
\end{equation}
which contains CSI, timestamps, and PHY metadata, Stage~(1) produces a calibrated and temporally thinned sequence with a canonical subcarrier layout:
\begin{equation}
\mathbf{X} = \{(\mathbf{x}^{(q)}, t_q, \mathbf{m}^{(q)})\}_{q=1}^{T'}, \qquad T' \le T,
\end{equation}
where $\mathbf{m}^{(q)}$ marks absent subcarriers due to bandwidth or waveform variations. Stage~(2) then embeds $\mathbf{X}$ into a fixed-length latent representation using attention over continuous time, enabling FuseFi to learn from CSI directly at the timestamps it is observed, without resampling or interpolation.
By separating waveform-related sanitization from time-aware learning, FuseFi preserves motion-induced dynamics in CSI while removing distortions introduced by opportunistic packet traffic and heterogeneous Wi-Fi protocols. Finally, the system generates the output for the sensing task.

\subsection{CSI Sanitization Pipeline}
\label{subsec:sanitization}

CSI extracted from normal communication packets carries useful motion-induced variations, but is also corrupted by packet-level amplitude scaling, heterogeneous PHY formats, and bursty arrivals within channel coherence time. FuseFi resolves these issues through a dedicated sanitization pipeline that converts raw CSI into a calibrated and compact representation suitable for our time-aware DNN.
\rev{Throughout the pipeline we operate on CSI amplitude. Raw CSI phase from COTS Wi-Fi devices is sensitive to carrier-frequency offset, sampling-frequency offset, packet-detection delay, and random per-packet phase offsets, especially under heterogeneous packet types, multiple bandwidths/bands, and independent sniffers. CSI amplitude is less sensitive to these impairments and can be stabilized through the normalization steps below; phase, Doppler, and multi-antenna features are discussed as future extensions in Section~\ref{sec:discussion}.}

{(i)~Clustering.}
Let $\tilde{\mathbf{H}}^{(p)}$ be the raw CSI of packet $p$.
Different PHY formats and bandwidths produce heterogeneous CSI waveforms. Therefore, FuseFi first groups packets using PHY metadata (\eg frame type, bandwidth).

{(ii)~Normalization.}
Packets may exhibit arbitrary amplitude scaling due to AGC and transmit-power adaptation, obscuring motion-induced changes.
Such amplitude inconsistency is non-trivial. Recent concurrent work has shown that AGC-related receiver effects can induce substantial CSI variability and significantly affect sensing robustness across different \textit{devices} under a highly controlled common-input setup~\cite{portner2026same}. In contrast, our setting reveals similar inconsistency across real heterogeneous \textit{communication packets} within a single link.
FuseFi stabilizes amplitude statistics by applying $L_2$ normalization across subcarriers of the CSI $\tilde{\mathbf{H}}^{(p)}$ to obtain the normalized CSI $\mathbf{H}^{(p)}_{\mathrm{norm}}$.
\rev{This step removes the per-packet global amplitude scale caused by AGC and transmit-power control, but preserves the normalized frequency-selective shape across subcarriers and its temporal evolution across packets.}
To prune waveform outliers within each group formed in step (i), we utilize the magnitude vector $\overline{\mathbf{a}}^{(p)}$ of this normalized CSI.
For a cluster ${\mathcal{C}}$, define the mean normalized magnitude as $\boldsymbol{\mu}$
and packets with Euclidean distance
\begin{equation}
d_p=\big\|\overline{\mathbf{a}}^{(p)}-\boldsymbol{\mu}\big\|_2 > \tau_d
\end{equation}
are discarded. This yields a set of \textit{clean clusters} $\{{\mathcal{C}}_c\}$, each with consistent waveform characteristics.

{(iii)~Alignment.}
Even after normalization, CSI from different clusters may still exhibit residual packet-dependent amplitude mismatch on overlapping subcarriers, due to imperfect gain removal and differences in PHY/LTF structures. FuseFi therefore performs alignment only between clusters that share compatible PHY/LTF structures.
Let $\mathcal{C}_c$ (source) and $\mathcal{C}_{c'}$ (reference) share overlapping subcarriers $\mathcal{K}_\cap$. Within a \emph{cross-cluster alignment window} $T_c$~\cite{xiong2013securearray,he2023sencom}, we form matched packet pairs
\rev{Here, $T_c$ is a design parameter for selecting approximately co-temporal packets across clusters for gain calibration, not the physical channel coherence time.}
\begin{equation}
\mathcal{P}=\{(p,q): p\!\in\!\mathcal{C}_c,\; q\!\in\!\mathcal{C}_{c'},\; |t_p-t_q|\le T_c\}.
\end{equation}
For each $(p,q)\!\in\!\mathcal{P}$, compute amplitude ratios
\begin{equation}
r^{(p,q)}_{k} = \frac{\overline{a}^{(p)}_{k}}{\overline{a}^{(q)}_{k}},\qquad k\in\mathcal{K}_\cap.
\end{equation}
The per-subcarrier scaling vector is the mean ratio $\boldsymbol{\gamma}$ over these pairs. 
Rescaling $\overline{a}^{(p)}_k / \gamma_k$ aligns cluster $\mathcal{C}_c$ to $\mathcal{C}_{c'}$. Clusters without compatible preambles or overlapping subcarriers remain separate channels.

{(iv)~Bursty Signal Filtering.}
Packet transmissions often arrive in micro-bursts, producing redundant CSI with negligible changes. To remove bias toward dense traffic while retaining motion information, FuseFi thins packets in each cluster $\mathcal{C}_c$ using a burst threshold $T_b$:
\begin{equation}
\Delta t_p = t_p - t_{p-1}, \qquad
\text{keep packet } p \text{ if } \Delta t_p \ge T_b.
\end{equation}
It reduces the sequence length from $T$ to $T' \!\le\! T$ without losing temporal diversity.

{(v)~Individual Subcarrier Selection (ISS) (Optional).}
Wideband CSI contains redundant subcarriers whose motion responses are negligible~\cite{zhang2019widetect,zeng2020multisense}, so retaining all of them not only increases computation but can also degrade learning performance. To reduce computation without sacrificing sensing quality, FuseFi optionally selects $K_{\mathrm{sel}}$ informative subcarriers per band. We compute the motion statistic $\mathrm{MS}_k$ for each subcarrier $k$ as the first-lag autocorrelation of its power response~\cite{zhang2019widetect}. We then uniformly partition the band into $K_{\mathrm{sel}}$ sub-bands and choose, in each sub-band $b$, the tone with maximum $\mathrm{MS}_k$:
\begin{equation}
\mathcal{S}_{\mathrm{ISS}}
= \big\{\arg\max_{k\in\mathcal{B}_b} \mathrm{MS}_k \,\big|\, b=1,\ldots,K_{\mathrm{sel}}\big\}.
\end{equation}
The sanitized input retains only these tones, with updated masks. Note that ISS is disabled for narrowband configurations (\eg 20\,MHz), where the subcarrier budget is already tight and pruning risks removing useful signal. The impact of ISS on accuracy and efficiency is validated in Section~\ref{subsubsec:efficiency}.

\subsection{Time-Aware DNN}
\label{subsec:fusefi_dnn}

After sanitization, the remaining CSI packets are informative but still arrive at non-uniform timestamps and vary in quantity across windows. Simply resampling them to a fixed rate would distort motion dynamics and introduce artificial transitions. FuseFi therefore processes CSI and their timestamps directly, using a time-aware attention architecture that aggregates irregular samples into a fixed-length representation.

\rev{FuseFi keeps the core mTAN mechanism: continuous-time embeddings map irregular observations onto a shared reference-time grid. We make two CSI-specific changes. First, we remove the missingness mask from learned attention features. Unlike clinical time series where missingness may be informative~\cite{che2018recurrent,shukla2021multi}, CSI missingness mainly reflects traffic opportunism, PHY format, and bandwidth availability rather than the activity itself; using it as a learned feature can therefore introduce a confound. FuseFi uses the mask only to indicate valid CSI positions during preprocessing/projection. Second, we form \emph{content-aware keys} by combining CSI content with timestamp embeddings, while queries depend only on reference time and values carry CSI content. This allows attention to depend on both what a packet measures and when it arrives.}

\noindent
\textbf{Inputs.} After sanitization, a window yields an irregular CSI set $\mathbf{X} = \{(\mathbf{x}^{(q)}, t_q, \mathbf{m}^{(q)})\}_{q=1}^{T'}$. We also specify a fixed query grid of $Q$ reference times
\begin{equation}
\mathcal{T}_{\text{ref}} = \{\tau_1,\ldots,\tau_Q\}
\end{equation}
within the same window.

\noindent
\textbf{ValueEmbedding.}
A lightweight encoder maps each instance to a feature vector
\begin{equation}
\mathbf{h}^{(q)} = f_{\text{enc}}(\mathbf{x}^{(q)}, \mathbf{m}^{(q)}) \in \mathbb{R}^{d_h}.
\end{equation}
Here, $\mathbf{m}^{(q)}$ is used to indicate available subcarrier positions when projecting heterogeneous-bandwidth CSI, and is not used as a learned feature in attention.

\noindent
\textbf{TimeEmbedding.}
To represent \emph{continuous} time, FuseFi embeds timestamps using learnable sinusoidal and linear functions:
\begin{equation*}
\boldsymbol{\phi}_h(t)[i] =
\begin{cases}
\omega_{0h} t + \alpha_{0h}, & i=0,\\[2pt]
\sin(\omega_{ih} t + \alpha_{ih}), & 0<i<d_r,
\end{cases}
\label{eq:unifi-time-emb}
\end{equation*}
with frequencies and phases $\{\omega_{ih},\alpha_{ih}\}$ learned during training. This generalizes transformer positional encodings to arbitrary real-valued time and enables attention queries at any reference $\tau_q$.

\noindent
\textbf{CombinedEmbedding.}
Unlike vanilla mTAN, which uses only time to construct keys, FuseFi forms \emph{content-aware keys} by fusing CSI features with their time features:
\begin{align}
\mathbf{e}_h^{(p)} &= \mathbf{W}_h\,\mathbf{h}^{(p)}, \qquad
\mathbf{e}_t^{(p)} = \mathbf{U}\,\boldsymbol{\phi}(t_p), \\[-3pt]
\mathbf{c}^{(p)} &= \mathbf{e}_h^{(p)} + \mathbf{e}_t^{(p)}.
\end{align}
The keys and values are then
\begin{align}
\mathbf{k}_p = \mathbf{W}_k\,\mathbf{c}^{(p)}, \qquad
\mathbf{v}_p = \mathbf{W}_v\,\mathbf{e}_h^{(p)},
\end{align}
while queries depend only on the reference time,
\begin{equation}
\mathbf{q}_q = \mathbf{W}_q\,\boldsymbol{\phi}(\tau_q).
\end{equation}

\noindent
\textbf{Attention to a fixed query grid.}
For each reference time $\tau_q$, FuseFi attends to the irregular CSI observations:
\begin{align}
\alpha_{q,p} &= \operatorname{softmax}_{p}\!\left(\frac{\mathbf{q}_q^\top \mathbf{k}_p}{\sqrt{d_k}}\right),\\
\mathbf{u}_q &= \sum_{p=1}^{T'} \alpha_{q,p}\,\mathbf{v}_p \in \mathbb{R}^{d_v}.
\end{align}
Collecting $\{\mathbf{u}_q\}_{q=1}^{Q}$ yields a \emph{fixed-length} matrix $\mathbf{U}\in\mathbb{R}^{Q\times d_v}$, regardless of how many packets arrived.

\noindent
\noindent
\textbf{Multi-band fusion.} FuseFi treats CSI from different frequency bands (e.g., 2.4~GHz and 5~GHz) as separate channels, each with its own timestamps and irregular sampling pattern.
Tokens from all bands are jointly attended against the reference query grid, producing a fused $\mathbf{u}_q$ at each reference time.
Because commodity single-radio sniffers (e.g., Raspberry Pi with Nexmon) can only capture one band at a time, per-band streams are collected by separate sniffers loosely synchronized via NTP.
Unlike bandwidth-synthesis approaches~\cite{xiong2015tonetrack,chen2019m,pegoraro2024hisac} that stitch sub-bands into a coherent wideband channel estimate and require precise phase alignment, FuseFi treats each band as an independent stream of timestamped amplitude observations.
\rev{FuseFi therefore does not require phase-coherent cross-band combining. Residual clock offsets only perturb when an amplitude token is placed in the continuous-time sequence, rather than corrupting the CSI amplitude itself, and the continuous-time embedding $\phi(t)$ absorbs such jitter as part of the irregular-sampling regime it already models. We empirically validate this timing-skew tolerance in Section~\ref{sec:evaluation}.}

\noindent
\textbf{No mask as feature.}
\rev{As discussed above, FuseFi uses the mask only to indicate valid CSI positions during preprocessing/projection, and excludes it from learned attention features to avoid traffic- or PHY-induced missingness confounds.}

\noindent
\textbf{Outputs.}
A GRU encoder aggregates the features into a compact representation, and then an MLP with softmax produces the final predictions $\hat{\mathbf{y}}$ for classification tasks, trained using cross-entropy loss.
Although we focus on classification tasks, the same backbone can be paired with a VAE-style decoder for CSI reconstruction, as shown in Section~\ref{subsubsec: interpolation}.

\section{Evaluation}
\label{sec:evaluation}

This section presents the real-world implementation and extensive evaluations of \ourSystem on five sensing tasks.

\subsection{Datasets and Tasks}
\label{subsec:datasets}

\begin{table*}[h!]
\centering
\caption{Dataset summary. Miss Rate (MR) vs 100\,Hz grid; Sampling Coefficient of Variation (SCV).}

\label{tab:datasets}

\begin{tabular}{llccccccc}
\toprule
Dataset 
& Task(s) 
& \makecell{Comm.\\CSI} 
& \makecell{Duration\\(min)} 
& \makecell{Band\\(GHz)} 
& \makecell{BW\\(MHz)} 
& \makecell{Rate\\(Hz)} 
& \makecell{Timestamp \\MR, SCV} 
\\
\midrule
WiGesture~\cite{zhao2024finding} & gestures & $\times$ & 48 & 2.4 & 20 & 100  & 0.09, 0.30 \\
WiFall~\cite{zhao2024knn} & actions, falls & $\times$ & 45 & 2.4 & 20 & 100 & 0.15, 0.33 \\
WiCount~\cite{zhao2024mining} & people count & $\times$ & 15 & 2.4 & 20 & 100 & 0.12, 0.28 \\
Injected-Fixed-Rate (Ref.)$^{*}$ & activities & $\times$ & 96 & 2.4, 5 & 20, 80 & 100  
& 0$^{*}$, 0$^{*}$ \\
\textbf{CommCSI-HAR} & activities & {$\checkmark$} & 96 & 2.4, 5 & 20, 20/80$^{+}$ & \textit{irregular}  & \textit{0.60/0.56}$^{\dagger}$, \textit{1.68/1.83}$^{\dagger}$ \\
 
\bottomrule
\end{tabular}

{\raggedright
$^{*}$ Reference (Ref.) datasets are collected \textit{simultaneously} with CommCSI-HAR while subjects performed activities at a fixed rate and are resampled to 100~Hz. 
$^{+}$ The bandwidth adapts between 20 and 80 MHz based on the communication needs. 
$^{\dagger}$ These metrics are calculated using timestamps from all frames, and for detailed metrics for each type, please refer to Table~\ref{table:csi_metrics}. 
\par}

\end{table*}

\subsubsection{\textbf{CommCSI-HAR Dataset}}

While many Wi-Fi sensing datasets exist, such as UT-HAR~\cite{yousefi2017survey}, Widar~\cite{zhang2021widar3}, SHARP~\cite{meneghello2022sharp}, NTU-Fi~\cite{yang2023sensefi}, Exposing the CSI~\cite{cominelli2023exposing}, XRF55~\cite{wang2024xrf55}, mmFi~\cite{yang2023mm}, OctoNet~\cite{yuanoctonet}, and CSI-Bench~\cite{zhu2025csi}, they all collect CSI from injected packets at high fixed rates with uniform frame types and bandwidths, as illustrated in Fig.~\ref{fig:challenges}(a). \textbf{No public dataset captures CSI from normal \emph{communication} traffic }with the heterogeneous waveforms, variable bandwidths, and irregular timestamps that characterize real ISAC scenarios. To fill this gap, we collect {CommCSI-HAR}, the first HAR dataset with irregularly sampled CSI from diverse communication packets across dual bands (2.4/5\,GHz) using COTS devices\footnote{The study was approved by our institution’s internal review board (IRB).}.

\begin{figure}[t]
    \centering
    \includegraphics[width=0.65\linewidth]{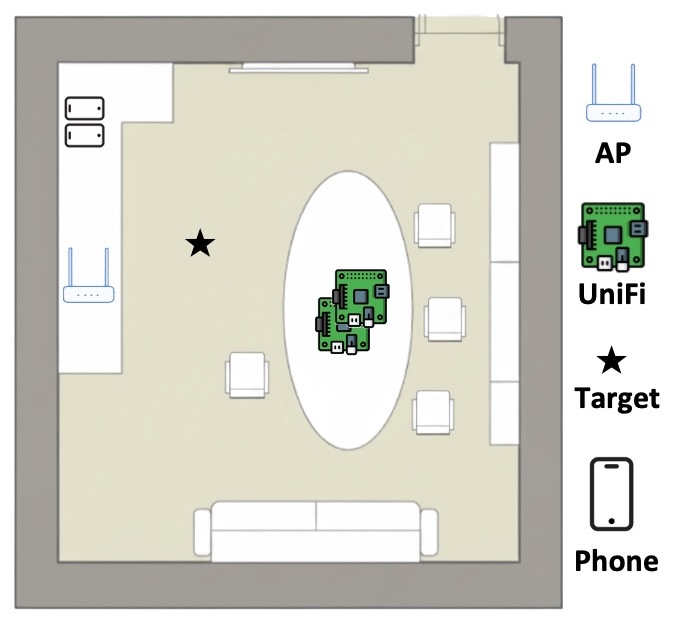}
    \caption{Data collection environment ($\sim$4m$\times$4m).}
    \label{fig:testbed}
\end{figure}

\paragraph{Real-World Collection Setup}
As shown in Fig.~\ref{fig:testbed}, we built a testbed in an office to capture CSI generated by normal communication traffic.
An ASUS dual-band AP (2.4/5\,GHz) \rev{supporting 802.11b/g/n/ac, with 802.11b disabled during collection because DSSS does not provide per-subcarrier OFDM CSI,} serves two smartphones streaming online video to produce communication traffic. 
\rev{Two Raspberry Pi 3B+ sniffers (one per band) with Broadcom BCM43455c0 Wi-Fi chipsets run the Nexmon tool~\cite{schulz2018nexmon} to capture CSI from downlink communication packets using legacy, HT, and VHT OFDM formats.}
On 5\,GHz we monitor an 80\,MHz channel, and the effective bandwidth adapts between 20 and 80\,MHz according to the communication needs.
\rev{The collected CSI therefore comes from OFDM-based 802.11g/n/ac traffic. FuseFi only requires per-packet OFDM CSI exposed by the receiver and is not tied to this specific AP generation.}
\rev{The Raspberry Pi sniffers are used only as prototype measurement instruments for CSI readout. They do not transmit sensing packets and therefore add no airtime or co-channel interference to the monitored channel.} 
\rev{The sniffers are synchronized via NTP, providing millisecond-level timestamp alignment.}

Eight subjects (4 female, 4 male) perform six daily activities: \textit{walking}, \textit{running}, \textit{boxing}, \textit{waving-hands}, \textit{sitting}, and \textit{circling-arm}. 
\rev{For each subject and activity, we record 2-min data and segment it into non-overlapping 4-second windows, with a stride of 4\,s and 0\% overlap. Thus, each recording produces 30 segments, and CommCSI-HAR contains $8\,\text{subjects}\times6\,\text{activities}\times30\,\text{windows}=1{,}440$ total segments. Each segment retains the native irregular packet timestamps within its 4-second window.}
\rev{We conducted these experiments in a 4\,m\,$\times$\,4\,m conference-table sensing area within a larger approximately 10\,m\,$\times$\,15\,m indoor office space, with fixed AP placement and sensing positions. This room-scale setting is comparable to common indoor CSI-HAR collection setups used in prior work~\cite{portner2026same,zhao2024mining,yang2023sensefi}. This single-indoor-environment protocol isolates the effects of communication-driven CSI heterogeneity, which is the primary focus of this paper; cross-environment generalization is discussed in Section~\ref{sec:discussion}.}

\paragraph{Baseline-compatible Subset}
To enable fair comparison with CSI-BERT2 and other baselines originally designed for single-type fixed-format CSI inputs, we construct a baseline-compatible subset of CommCSI-HAR, denoted as CommCSI-HAR$^{*}$. This subset consists of 52 subcarriers extracted from QoS data frames in the 5~GHz band, matching the input format of CSI-BERT2 with minimal modification to the baseline architectures. 

\paragraph{Controlled Reference Setup}
For a controlled \textit{\textbf{reference}}, two additional Raspberry Pis inject dummy QoS data packets at around $150$\,Hz on both 2.4 and 5\,GHz {on channels \textit{distinct} from those used for CommCSI-HAR}, while another two Raspberry Pi sniffers \textit{\textbf{simultaneously}} capture the injected CSI, which are then downsampled to 100\,Hz.
We treat this Injected-Fixed-Rate (IFR) dataset as a reference setting collected under conventional fixed-rate injection.
\rev{The IFR reference is collected on separate channels and separate sniffers to avoid adding airtime, contention, or interference to the ambient communication traffic; therefore, it should be interpreted as a fixed-rate reference rather than a channel-matched ablation.}

\paragraph{Synthetic Datasets with Controlled Irregularity and Sparsity}
In addition to the real communication dataset, we construct synthetic variants of the Injected-Fixed-Rate stream to evaluate the robustness and operating limits of \ourSystem under a broad spectrum of communication conditions. Specifically, we generate synthetic datasets by subsampling the fixed-rate 20~MHz Injected-Fixed-Rate stream to control \emph{sparsity} through the missing rate (MR $\in [0,0.9]$) and \emph{irregularity} through the sampling coefficient of variation (SCV $\in [0,3]$). Here, MR$=0$ and SCV$=0$ correspond to the original fixed-rate data, whereas MR$=0.9$ and SCV$=3$ represent extremely sparse packet arrivals and highly irregular packet intervals. By varying packet availability and temporal irregularity in this way, these synthetic datasets span conditions ranging from dense and nearly regular traffic to highly sparse and bursty arrivals, thereby enabling a controlled study of sensing robustness under varying communication workloads.

\subsubsection{\textbf{More Datasets and Tasks} (fixed-rate injection with minor packet loss)}
In addition, we evaluate \ourSystem on four tasks using three public datasets. 
They extract CSI from \textit{\textbf{injected}} packets at a target rate of $100$\,Hz with \textit{minor packet loss}, using an ESP32 receiver over a 20\,MHz channel in the 2.4\,GHz band.

Table~\ref{tab:datasets} summarizes all datasets and tasks, and CommCSI-HAR provides the first CSI dataset extracted from normal communication traffic, exhibiting heterogeneous waveforms, variable bandwidths, and irregular/bursty timestamps. 
To quantify timestamp characteristics, we introduce two metrics: the \textit{{Missing Rate (MR)}} and the \textit{{Sampling Coefficient of Variation (SCV)}}. MR measures \textit{sparsity} as the percentage of fixed-duration time bins that contain no data, while SCV captures \textit{irregularity} as the ratio of the standard deviation to the mean of inter-packet intervals. Both metrics of these public datasets are much smaller compared with our dataset, consistent with the fixed-rate nature of injected packets.

\begin{table*}[h!]
\centering
\caption{Dataset quality: accuracy of the FuseFi-DNN for different datasets. \textbf{Bold} = best, \underline{underline} = second-best. }

\label{table:nodes_all}
\begin{tabular}{llcc}

\toprule
\textbf{Dataset} & \textbf{Band(s)} & \textbf{Packet Type} & \textbf{Accuracy (mean $\pm$ std)} \\
\midrule
\multirow{4}{*}{\textbf{CommCSI-HAR}} 
  & 5 GHz        & QoS Data & .9143 $\pm$ .0127 \\
  \cline{2-4}
  & 5 GHz        & All & .9431 $\pm$ .0104 \\
  & 2.4 GHz      & All  & .8611 $\pm$ .0128 \\
  & 5 $+$ 2.4\,GHz & All         & \textbf{.9688} $\pm$ .0054 \\
\midrule
\multirow{3}{*}{CommCSI-HAR $+$ Linear Interpolation} 
  & 5\,GHz        & All & .9319 $\pm$ .0071 \\
  & 2.4\,GHz      & All & .8079 $\pm$ .0102 \\
  & 5 $+$ 2.4\,GHz & All         & .9229 $\pm$ .0102 \\
\midrule
\multirow{3}{*}{Injected-Fixed-Rate (For Reference)} 
  & 5\,GHz        & -                          & .9310 $\pm$ .0089 \\
  & 2.4\,GHz      & -                          & .8690 $\pm$ .0166 \\
  & 5 $+$ 2.4\,GHz & -         & \underline{.9465} $\pm$ .0084 \\
\bottomrule
\end{tabular}

{\raggedright

\par}

\end{table*}

\begin{figure*}[t]
    \centering
    \includegraphics[width=0.9\linewidth]{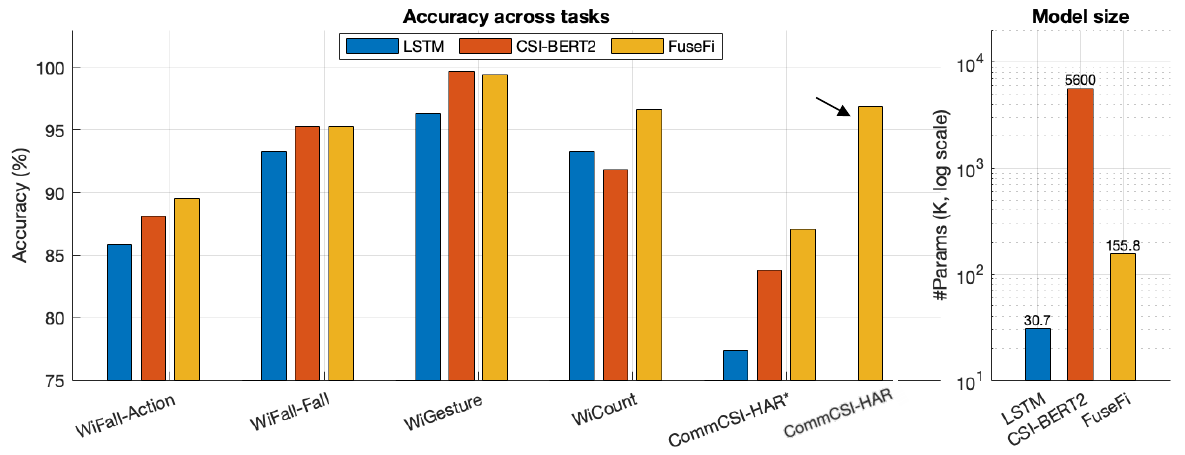}
    \caption{Model capacity across diverse tasks: accuracy and size of FuseFi-DNN and the baselines. 
    CommCSI-HAR$^{*}$ is the baseline-compatible subset of the CommCSI-HAR dataset. 
    The arrow highlights the additional 8.8\% accuracy gain obtained when using the full CommCSI-HAR dataset.}
    \label{fig:more_baselines_tasks}
\end{figure*}

\subsection{Implementation Details}
\label{subsec:impl}

All CSI data are sanitized in MATLAB R2021b. We set the distance threshold  $\tau_d$ to 0.6, the burst threshold $T_b$ to 10ms, and the cross-cluster alignment window $T_c$ to 1ms. 
\rev{Here, $T_c$ is a cross-cluster alignment window used to match near-coincident packets for gain calibration. We set $T_c=1$\,ms to be well below the tens-of-milliseconds physical coherence timescale of human-motion Wi-Fi channels at 2.4/5\,GHz.}
These values are practical defaults chosen to balance temporal fidelity, computational efficiency, and the coherence properties of commodity Wi-Fi links; adaptive thresholding is a natural extension but is not required for the gains reported here. 
\rev{A sensitivity sweep in Section~\ref{subsubsec:threshold_sensitivity} shows that FuseFi is stable across practical choices of $\tau_d$, $T_b$, and $T_c$.}
Model training and inference of FuseFi DNN are performed on a server with an AMD EPYC~9354 CPU and an NVIDIA H100 GPU, running Python~3.12.1, CUDA~11.8, and PyTorch~2.2.1. The embedding sizes are set to 64, and the reference point number is 64.
We use the Adam optimizer with a learning rate 0.001 and a batch size $64$.
Each sample corresponds to a sequence extracted from a $4\,\mathrm{s}$ time window.
\rev{The train/test split is $80/20$, and the 80\% training portion is further split 80/20 into training and validation sets.}
Results are averaged over $5$ random seeds.
\rev{This is a random, subject-mixed segment-level split, in which segments from all subjects are pooled before partitioning, following the common protocol used in related Wi-Fi-CSI HAR work~\cite{yang2023sensefi,zhao2024mining}.}
We evaluate models using accuracy.

\subsection{Baselines}
\label{subsec:baselines}

To assess dataset quality, we utilize the simultaneously collected Injected-Fixed-Rate reference dataset as a fixed-rate baseline for comparison.
Additionally, we evaluate linear interpolation to show its impact on CSI quality, as it's a widely used preprocessing step in Wi-Fi sensing~\cite{yang2022wiimg}.

To demonstrate the capabilities of \ourSystem-DNN, we benchmark it against LSTM~\cite{hochreiter1997long}, CSI-BERT2~\cite{zhao2024mining}, and three additional irregular-time / CSI-HAR baselines described below. LSTM serves as a fundamental baseline capable of processing varying inputs; CSI-BERT2 represents a strong baseline employing a BERT-based framework for robust sensing under lossy CSI conditions.
\rev{We re-evaluated CSI-BERT2 using the authors' official open-source implementation and evaluation protocol, and report the mean and standard deviation over five random seeds. Across the four public-benchmark tasks, our CSI-BERT2 results differ from the published numbers in~\cite{zhao2024mining} by at most 2.5 percentage points, indicating that the comparison is reproducible.}

\rev{\textbf{Additional irregular-time and CSI-HAR baselines.}
To strengthen the comparison on irregular communication CSI, we additionally evaluate GRU-D~\cite{che2018recurrent}, a decay-based irregular-time RNN, GRU-ODE-Bayes~\cite{brunzema2022gru}, a continuous-time irregular-time model, and WiGRUNT~\cite{gu2022wigrunt}, a ResNet-based CSI activity-recognition model. All three baselines are evaluated on the same baseline-compatible CommCSI-HAR$^{*}$ used for the LSTM and CSI-BERT2 comparisons.}

\rev{As shown in Table~\ref{tab:irregular_baselines}, these additional baselines achieve competitive performance under the matched 20-MHz setting. GRU-D reaches 77.7\%, GRU-ODE-Bayes reaches 82.3\%, and WiGRUNT reaches 75.3\% accuracy, whereas \ourSystem{}-DNN reaches 88.1\%. Among the newly added baselines, GRU-ODE-Bayes performs best, indicating that continuous-time modeling is effective for irregular communication CSI. Nevertheless, \ourSystem{}-DNN retains a 5.8-percentage-point advantage under the same evaluation protocol. These results show that while stronger generic irregular-time models and a larger ResNet-based CSI-HAR architecture substantially improve performance, the proposed CSI-specific time-aware architecture remains the most effective under the matched setting.}

\begin{table}[t]
\centering
\caption{\rev{Additional irregular-time and CSI-HAR baselines.}}
\label{tab:irregular_baselines}
\resizebox{\columnwidth}{!}{%
\begin{tabular}{lcc}
\toprule
\textbf{Model} & \textbf{\#Params} & \textbf{Accuracy} \\
\midrule
GRU-D~\cite{che2018recurrent}              & 43.4K   & 77.7\% \\
GRU-ODE-Bayes~\cite{brunzema2022gru}       & 291.9K  & 82.3\% \\
WiGRUNT~\cite{gu2022wigrunt} (2-D ResNet)  & 11.17M  & 75.3\% \\
\midrule
\ourSystem{}-DNN                           & 155.8K  & \textbf{88.1\%} \\
\bottomrule
\end{tabular}
}
\end{table}

\rev{\textbf{Passive beacon-only baseline.}
To provide a passive-sensing baseline under the same no-packet-injection paradigm as \ourSystem{}, we reprocess the traffic captured in CommCSI-HAR and retain only beacon-frame CSI. The AP broadcasts these beacons as part of normal Wi-Fi operation, so no sensing-specific probe packets are injected. The resulting beacon-only dataset provides CSI at approximately 10\,Hz. We evaluate both \ourSystem{}-DNN and a conventional LSTM on this beacon-only dataset using the same training and evaluation protocol as in Table~IV.}

\rev{On the beacon-only dataset, \ourSystem{}-DNN and LSTM achieve $84.35\%$ and $81.10\%$ accuracy, respectively. Their relatively modest performance difference is consistent with the approximately periodic and sparse nature of beacon traffic, for which irregular-time modeling is less consequential. In contrast, \ourSystem{} achieves $96.88\%$ on the full naturally occurring traffic, improving accuracy by 12.53 percentage points over the beacon-only \ourSystem{}-DNN baseline. This result shows that exploiting the full, diverse naturally occurring traffic substantially outperforms relying only on sparse, periodic beacons.}

Direct comparison with SenCom~\cite{he2023sencom} is infeasible due to fundamental methodological differences. Unlike SenCom, which relies on active packet injection, \ourSystem exploits diverse ambient traffic from multiple bands to preserve communication throughput. These distinct paradigms result in disparate traffic conditions and evaluation setups, preventing exact numerical comparisons across the two systems.

\subsection{Overall Performance}
\label{subsec:overall_perf}

\subsubsection{\textbf{CommCSI-HAR Dataset Quality}}
\label{subsec:results_perf}

Table~\ref{table:nodes_all} reports FuseFi’s accuracy under three data regimes, all using the same FuseFi-DNN model and training protocol: (i) \textbf{CommCSI-HAR}, (ii) \textbf{CommCSI-HAR + Linear Interpolation}, and (iii) \textbf{Injected-Fixed-Rate} (reference dataset collected under fixed-rate injection). Here are some key observations.

\noindent
\textbf{Impact of Non-QoS Data and Multi-Band Fusion.}
Evaluation on the CommCSI-HAR dataset demonstrates that incorporating diverse frame types within the 5~GHz band, or fusing data across bands, consistently outperforms relying solely on QoS data. This confirms that the diversity inherent in natural traffic, spanning various frame types, bandwidths, and frequencies, provides complementary sensing cues that single-format CSI may miss. These findings underscore the necessity of our framework, which is uniquely designed to sanitize and fuse these heterogeneous signals to maximize sensing performance.

\noindent
\textbf{Effectiveness on CommCSI-HAR.}
On our dataset, \ourSystem performs competitively with the Injected-Fixed-Rate dataset across all input choices: single-band 5/2.4~GHz, and dual-band fusion.
\rev{This comparison is intended as a competitiveness check rather than a channel-matched ablation. As noted in Section~\ref{sec:evaluation}, the IFR reference is collected on separate channels using separate sniffers to avoid contaminating the ambient traffic; therefore, part of the observed difference may reflect channel, sniffer, or interference differences rather than the sampling regime alone.}

\rev{To disentangle architectural gains from data-source gains, we compare models under matched conditions. On CommCSI-HAR$^{*}$, FuseFi-DNN reaches 88.1\%, compared with 77.8\% for LSTM and 83.8\% for CSI-BERT2, isolating the architectural gain on the same fixed-format input. Holding the FuseFi architecture fixed, expanding the input from 5\,GHz QoS-only CSI to all 5\,GHz packet/frame types improves accuracy to 94.3\%, showing the benefit of packet diversity and bandwidth. Adding 2.4\,GHz CSI for dual-band fusion further improves accuracy to 96.9\%, showing the benefit of multi-band sensing. These results indicate that FuseFi's overall gain comes from both its time-aware architecture on matched fixed-format CSI and its ability to exploit richer heterogeneous communication CSI across packet/frame types and frequency bands.}

\noindent
\textbf{Interpolation Hurts.}
Across all input settings, {CommCSI-HAR + Linear Interpolation} yields consistently lower accuracy than the original dataset. While interpolation produces a regular, fixed-rate sequence, it also injects artifacts unrelated to motion,
which degrades downstream accuracy.

\noindent
\rev{
\textbf{Absolute-Energy vs. Normalized CSI.}
Because L$_2$ normalization removes each packet's global amplitude scale, we evaluate whether explicit energy-variance descriptors recover useful information beyond the normalized CSI. We augment the L$_2$-normalized dual-band CommCSI-HAR input with two per-packet, per-band descriptors: magnitude variation coefficient (MVC) and normalized-power-spectrum spectral entropy. Under the same training protocol, the default L$_2$-normalized FuseFi reaches 96.88$\pm$0.54\%; adding MVC gives 97.22$\pm$0.31\%, adding spectral entropy gives 97.43$\pm$0.28\%, and adding both gives 97.15$\pm$0.51\%. These differences are within seed-level variation and do not show a consistent improvement. 
Because the aggregate accuracy is already near the performance ceiling, this ablation has limited resolution for revealing class-specific gains; under the evaluated aggregate metric, the energy-variance descriptors provide no consistent measurable benefit beyond the normalized CSI sequence.
In contrast, removing L$_2$ normalization drops accuracy to 92.71$\pm$0.65\%, indicating that raw absolute packet scale is affected by AGC and transmit-power variations in this heterogeneous communication-CSI setting.
}

\noindent
\textbf{5~GHz vs. 2.4~GHz Bands.}
Across all datasets, the 5~GHz band consistently outperforms 2.4~GHz. This aligns with expectations: 5~GHz supports wider bandwidths with higher spectral resolution. However, since 2.4~GHz offers superior signal penetration through obstacles, fusing both bands leverages these complementary strengths to consistently yield the highest accuracy.

\subsubsection{\textbf{FuseFi’s Performance on Five Sensing Tasks}}

Figure~\ref{fig:more_baselines_tasks} summarizes the performance of our work and the baselines. \textbf{\ourSystem achieves state-of-the-art accuracy on all tasks while requiring only $\sim$3\% parameters\footnote{\rev{The exact count varies slightly with the input configuration. For a fair comparison, we report the parameter counts of all models under the same input setting.}} of CSI-BERT2}. While LSTM offers the smallest model size, it performs significantly worse due to its inability to model irregular time intervals.
On public datasets, \ourSystem's gains are modest as the injected streams are near-regular.
On the CommCSI-HAR dataset, comparison is performed on a compatible subset for minimal modification to CSI-BERT2. \ourSystem demonstrates pronounced improvement here, reflecting its robustness to irregularity. Furthermore, utilizing the full CommCSI-HAR dataset yields an additional 8.8\% accuracy gain, demonstrating \ourSystem's adaptability to diverse and complex traffic.

\noindent
\rev{\textbf{Subject-Independent Generalization and Per-User Calibration.}
To address the limitation of the subject-mixed 80/20 split, we further evaluate leave-one-subject-out (LOSO) cross-validation and lightweight per-user calibration on CommCSI-HAR. Under LOSO, FuseFi drops from 96.9\% under the in-domain split to 41.6\%, confirming that the subject-mixed result should not be interpreted as subject-independent performance. This degradation is consistent with broader trends in Wi-Fi CSI-based HAR: FuseFi on WiFall-Action drops from 89.6\% under the in-domain protocol to 43.2\% under LOSO, while the CSI-BERT2 baseline drops from 88.1\% to 40.1\% on the same dataset under the corresponding protocols. 
We further evaluate a practical few-shot calibration setting. For each LOSO fold, we freeze the encoder trained on the seven source subjects and fit only a linear classification head using a small labeled support set from the held-out subject. The calibration and evaluation windows are strictly disjoint, and no encoder parameters are updated. Accuracy increases to 91.3\% with five labeled windows per activity and 94.7\% with ten labeled windows per activity. Since each window spans 4 seconds, these settings correspond to approximately 2 and 4 minutes of labeled calibration data, respectively. These results indicate that most of the in-domain performance can be recovered through lightweight user-specific calibration. 
We discuss the deployment implications and adaptation-based future directions in Section~\ref{subsubsec:env_gen}.
}

\begin{figure}[t]
    \centering
    \includegraphics[width=1.0\linewidth]{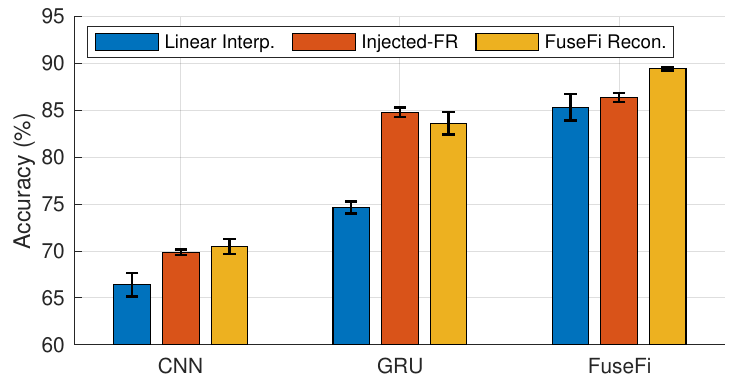}
    \caption{Accuracy of models evaluated on different fixed-rate CSI data: Linear Interp.\ (linearly interpolated data; widely used baseline), Injected-FR (injected fixed-rate data for reference), and FuseFi Recon.\ (ours). Here we use CommCSI-HAR$^{*}$ noted in Figure~\ref{fig:more_baselines_tasks}.}
    \label{fig:Restore_Compare}
\end{figure}

\subsection{Generalization and Robustness}

\subsubsection{\textbf{Fixed-Rate CSI Reconstruction}}
\label{subsubsec: interpolation}
As discussed in Section~\ref{subsec:fusefi_dnn}, the FuseFi backbone can be coupled with a VAE-style decoder to generate fixed-rate CSI sequences from irregular inputs, adopting the generative methodology of mTAN~\cite{shukla2021multi}. We define this process as ``reconstruction'' rather than ``interpolation'' because the decoder samples from a learned conditional distribution on a target grid, rather than merely filling temporal gaps. Figure~\ref{fig:Restore_Compare} demonstrates that \textit{FuseFi-reconstructed} CSI consistently outperforms linear interpolation across all evaluated models. Moreover, its accuracy aligns with the fixed-rate reference dataset, confirming that the reconstruction preserves sensing-relevant structure without introducing the artifacts.

\subsubsection{\textbf{Robustness to Irregularity and Sparsity}}
\label{subsubsec:robustness}
Using the synthetic datasets with controlled irregularity and sparsity, Figure~\ref{fig:Heatmap_MR_SCV} shows that \ourSystem maintains high accuracy for MR $\leq 0.5$ over a wide range of SCV values, and degrades gracefully as sparsity increases further. The effect of irregularity becomes more pronounced when sparsity is low. In practical deployment, however, \ourSystem alleviates this challenge by fusing diverse CSI from dual-band traffic, which effectively lowers both MR and SCV and allows the system to operate in a more favorable regime.

\rev{Here, MR measures sampling incompleteness, so $1-\mathrm{MR}$ corresponds to sampling completeness and high MR represents a low-traffic or low-observation regime. The MR/SCV sweep therefore directly evaluates sparse and irregular packet observations. Accuracy remains within 84.9--86.8\% for MR $\leq 0.5$ across the tested SCV range, and \ourSystem{} still reaches 85.2\% at MR $=0.9$ with regular timing. The main degradation appears only in the joint worst case where traffic is both extremely sparse and highly irregular (MR $=0.9$, SCV $=3$, 73.7\%).}

\rev{The measured CommCSI-HAR operating point lies away from this worst-case corner: the per-band MR is approximately 0.56--0.60 and the per-band SCV is approximately 1.68--1.83. Cluster and band fusion further mitigate sparse packet co-occurrence in realistic traffic. Individual packet clusters have missing rates of 0.70--0.87, while fusing clusters within each band reduces the effective missing rate to 0.56 on 5\,GHz and 0.60 on 2.4\,GHz, with dual-band fusion providing additional coverage.}

\subsubsection{\textbf{\rev{Sensitivity to Sanitization Thresholds}}}
\label{subsubsec:threshold_sensitivity}
\rev{We further performed an end-to-end threshold-sensitivity sweep on the dual-band CommCSI-HAR configuration, retraining FuseFi while varying one threshold at a time around the default settings $\tau_d=0.6$, $T_b=10$\,ms, and $T_c=1$\,ms. Specifically, we tested $\tau_d\in\{0.2,0.3,0.4,0.5,0.6\}$, $T_b\in\{5,10,20\}$\,ms, and $T_c\in\{0.5,1,2,5,10\}$\,ms. Relative to their respective default settings, these variations changed the final accuracy by at most 0.5, 0.4, and 0.1 percentage points for $\tau_d$, $T_b$, and $T_c$, respectively. The more aggressive $\tau_d=0.1$ setting falls outside the usable range because it prunes entire CSI streams from many clips. Varying $T_c$ produced minimal change because $T_c$ only determines which near-coincident packet pairs are used to estimate the cross-cluster gain vector $\gamma$, and the resulting gain estimates remained stable. On the less redundant baseline-compatible subset, performance remains stable for $\tau_d\in[0.3,0.6]$, whereas the more aggressive $\tau_d=0.2$ causes an approximately 3-percentage-point decrease, suggesting greater sensitivity to aggressive pruning when input diversity and redundancy are limited. These results indicate that FuseFi is not fragile to reasonable sanitization-threshold choices within the evaluated ranges, rather than relying on finely tuned values.}

\subsubsection{\textbf{\rev{Cross-Band Timing-Skew Robustness}}}
\label{subsubsec:timing_skew}

\rev{To validate the millisecond-level NTP synchronization assumption, we perform a controlled synthetic-skew experiment on the dual-band configuration. We treat one sniffer as the time reference and inject an additional timestamp offset plus per-packet jitter into the other sniffer's timestamps, sweeping the skew level over 0, 5, 10, and 20\,ms. The no-skew case reproduces the dual-band operating point, and accuracy remains stable across the sweep: 96.88$\pm$0.54\%, 97.08$\pm$1.07\%, 97.50$\pm$0.60\%, and 97.43$\pm$0.78\%, respectively. The largest change from the no-skew condition is below one percentage point and within seed-to-seed variation, and accuracy does not degrade at any level; even the 20\,ms stress level---twice the 100\,Hz reference sampling interval and well above the single-digit-millisecond NTP regime---does not materially affect accuracy, indicating that \ourSystem{}'s amplitude-based cross-band fusion is not fragile to millisecond-level timing misalignment.}

\begin{figure}[t]
    \centering
    \includegraphics[width=1.0\linewidth]{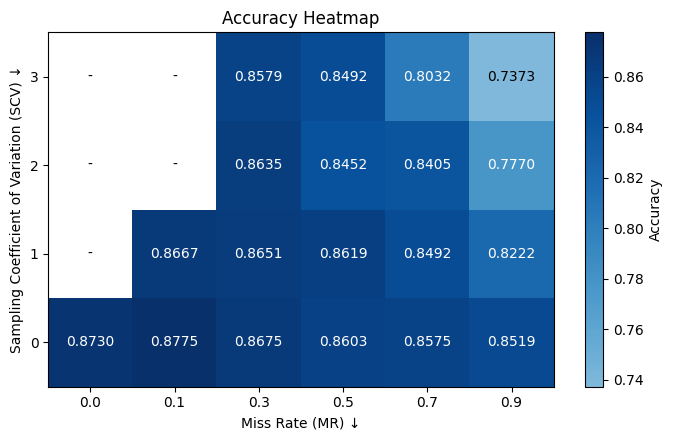}
    \caption{Accuracy heatmap for synthetic HAR dataset with different MR and SCV.
    }
    \label{fig:Heatmap_MR_SCV}
\end{figure}

\subsection{Ablation Study}

\subsubsection{\textbf{Effects of Sanitization on CSI Quality}}
\begin{figure}[t]
    \centering
    \includegraphics[width=1.0\linewidth, height=0.65\linewidth]{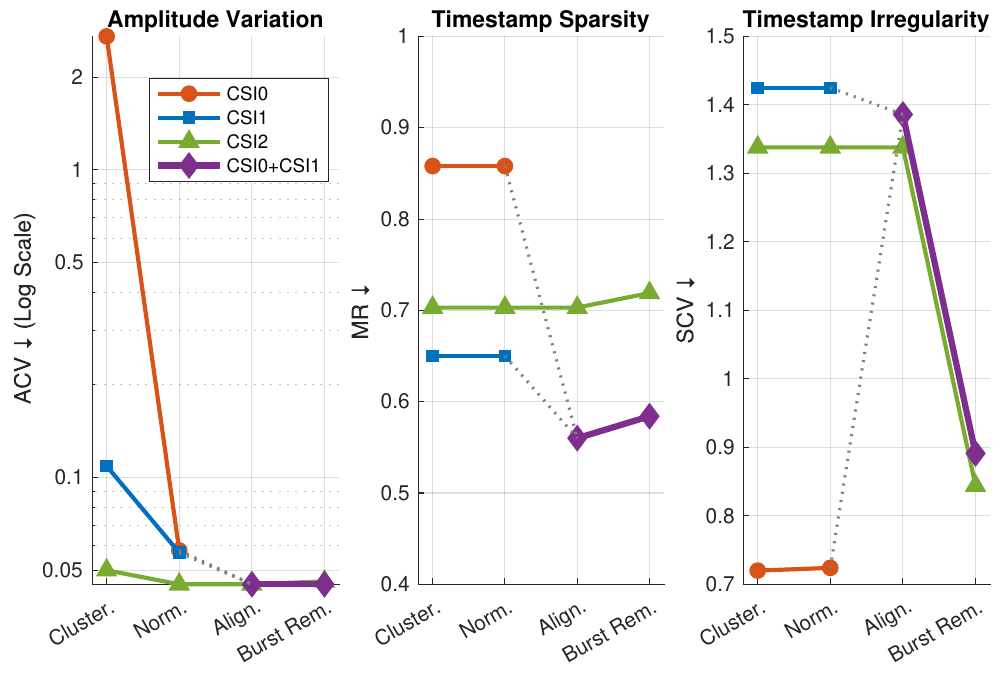}
    \caption{CSI Quality Across Sanitization Pipeline.}
    \label{fig:csi_quality}
\end{figure}

Figure~\ref{fig:csi_quality} illustrates the evolution of CSI quality throughout the sanitization pipeline, validating the necessity of each stage. In addition to the timestamp metrics (MR and SCV), we introduce the \textit{{Amplitude Coefficient of Variation (ACV)}}~\cite{hu2024you} to quantify amplitude stability. Calculated as the average coefficient of variation across all subcarriers in a static environment, the ACV serves as a proxy for signal quality, where higher values indicate increased instability due to power adaptations or environmental noise.
We utilize a 2-minute CSI sequence collected on a 20 MHz channel in the 5 GHz band. This data was acquired in a static environment, which is required for the ACV calculation~\cite{hu2024you} to show non-motion amplitude noise.
Following the \textit{Clustering} stage, three distinct clusters are identified, and we denote them as CSI0, CSI1, and CSI2. Subsequently, \textit{Normalization} substantially improves the amplitude quality. The \textit{Alignment} of CSI0 and CSI1 then improves their overall timestamp sparsity. Finally, the application of \textit{Burst Removal} significantly reduces timestamp irregularity, with minimal cost to MR. Note that we omit the ISS stage in this analysis, as motion statistics are inapplicable to static scenes. The impact of the pipeline, including ISS, on sensing accuracy or/and model efficiency is evaluated in following sections in the applicable scenarios.

\begin{figure}[t]
    \centering
    \includegraphics[width=1.0\linewidth]{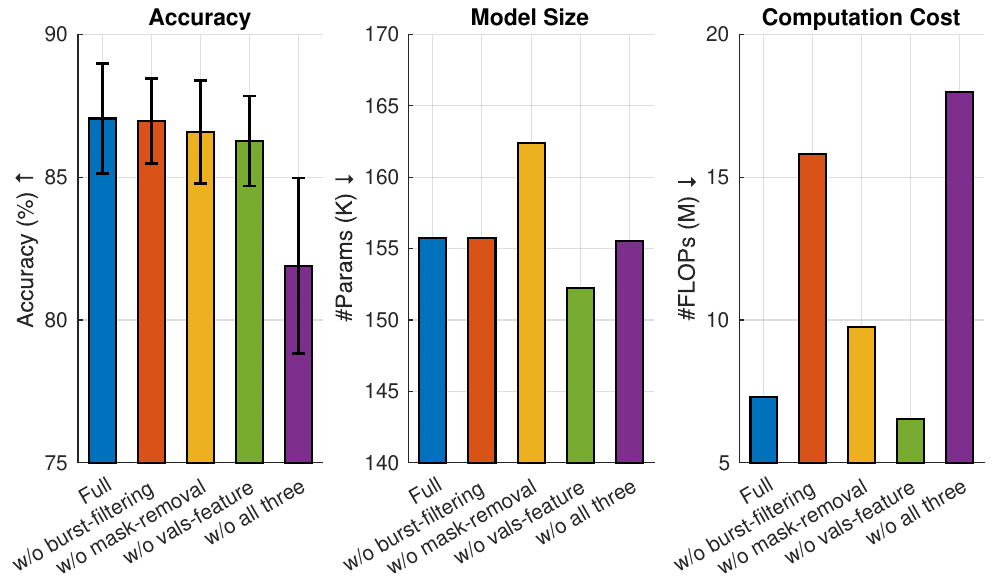}
    \caption{Ablation study: effects of DNN architecture and bursty CSI filtering.}
    \label{fig:ablation_dnn}
\end{figure}

\subsubsection{\textbf{Ablations: burst filtering and DNN design}}
\label{subsec:ablation}

Figure~\ref{fig:ablation_dnn} shows ablations on FuseFi design assessing (i) bursty CSI filtering in sanitization and (ii) two architectural choices in FuseFi\textendash DNN: removing \emph{mask-as-feature} inputs, and using \emph{content-aware keys} (CSI features) in attention. Here we use the same data denoted by $^{*}$ in Figure~\ref{fig:more_baselines_tasks}.
Full \ourSystem achieves the highest accuracy. Omitting burst filtering (\textit{w/o burst-filtering}) results in a minor accuracy drop but doubles the computation cost due to the processing of redundant CSI. Reintroducing mask features (\textit{w/o mask-removal}) further degrades accuracy while significantly increasing both model size and computational overhead. This confirms that unlike clinical time series where missingness is often informative, CSI sparsity stems primarily from PHY/traffic adaptations rather than motion; thus, explicitly modeling masks introduces noise and confounds learning. Finally, removing all components produces the most significant performance decline, demonstrating that our sanitization and architectural modules are complementary, each addressing distinct challenges of irregularly sampled CSI.

\subsubsection{\textbf{Impact of Individual Subcarrier Selection}}
\label{subsubsec:efficiency}

\begin{figure}[t]
    \centering
    \includegraphics[width=1.0\linewidth]{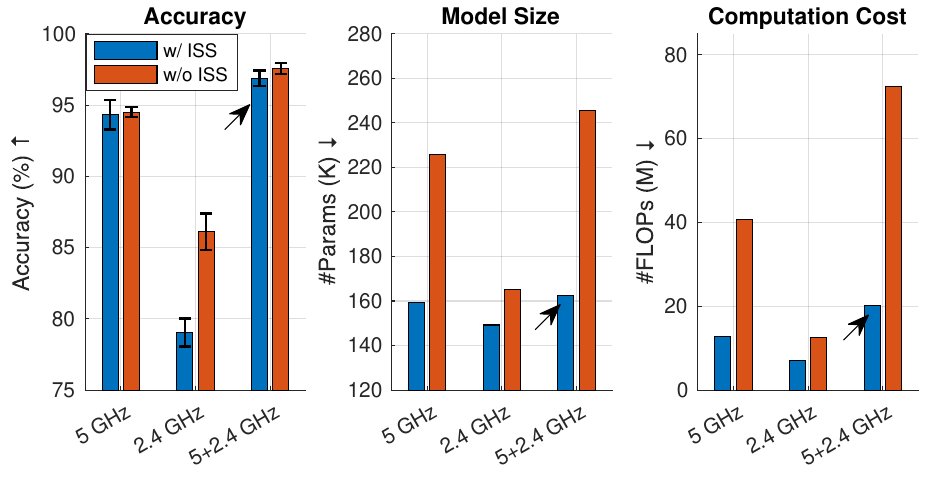}
    \caption{Impact of Individual Subcarrier Selection (ISS) on sensing performance and efficiency. The arrow indicates the configuration adopted in FuseFi.}
    \label{fig:effect_iss}
\end{figure}

Figure~\ref{fig:effect_iss} shows that enabling \emph{ISS} reduces model size and computation cost across different settings. On 5~GHz, ISS preserves accuracy while improving efficiency, consistent with the intuition that wide channels contain redundant tones. On 2.4~GHz, ISS can degrade sensing because the subcarrier budget is already tight and pruning removes useful signal; in such narrowband cases, ISS can be disabled. In the dual-band configuration, ISS maintains accuracy while lowering compute and latency, making it a sensible default for our efficient and robust design.

\subsubsection{\textbf{Fusion of CSI from different frame types within a single band}}

\begin{table}[t]
\centering
\caption{CSI Timestamp Metrics.}
\begin{tabular}{cccc}
    \toprule
    \textbf{Band} & \textbf{CSI Cluster$^{\dagger}$} & \textbf{MR} & \textbf{SCV} \\ 
    \midrule
    
    \multirow{4}{*}{5 GHz} 
        & CSI0 & 0.8592 & 0.7088 \\
        & CSI1 & 0.6970 & 1.8377\\
        & CSI2 & 0.7503 & 2.1390 \\
        & All  & 0.5593 & 1.8295\\
    \midrule

    \multirow{4}{*}{2.4 GHz} 
        & CSI0 & 0.8699 & 0.5639 \\
        & CSI1 & 0.7603 & 1.6300\\
        & CSI2 & 0.7909 & 2.0141 \\
        & All  & 0.6044 & 1.6832\\
    \bottomrule
\end{tabular}
\label{table:csi_metrics}

{\raggedright
$^{\dagger}$ Three CSI clusters discovered during sanitization. 
\par}

\end{table}

\label{subsubsec:framefusion_singleband}
Table~\ref{table:csi_metrics} reports the missing rate (MR) and sampling coefficient of variation (SCV) for the three CSI clusters discovered during the sanitization stage. 
CSI2 (QoS data frames) shows the largest SCV on both 5~GHz and 2.4~GHz, reflecting highly bursty traffic. 
In contrast, CSI0, including periodic management frames such as beacons, is much more regular (lower SCV) but suffers from higher sparsity (higher MR). 
When we \emph{combine all clusters} within a band, MR drops substantially, and SCV is also reduced relative to the burstiest stream. 
These statistics justify fusing heterogeneous CSI from diverse frame types within a band: periodic management frames provide temporal coverage that counters sparsity, while bursty data frames supply rich motion cues with wider bandwidths, together yielding a more balanced, informative sequence for sensing.

\begin{table}[t]
\centering
\caption{Accuracy of the CSI clusters in 5~GHz band.}
\resizebox{\columnwidth}{!}{%
\begin{tabular}{llcl}
    \toprule
    \textbf{CSI Cluster} & \textbf{Bandwidth (MHz)} & \textbf{Accuracy} \\ 
    \midrule
    
    \multirow{1}{*}{ CSI0} 
        & 20 & .8810 ± .0043  \\
    \midrule

    \multirow{2}{*}{ CSI1} 
        & 20 & .8571 ± .0305 \\
        & 80 & .9317 ± .0108 \\
    \midrule

    \multirow{2}{*}{ CSI2} 
        & 20 & .8706 ± .0192 \\
        & 80 & .9143 ± .0127 \\
    \midrule

    \multirow{2}{*}{ [CSI0,CSI1]} 
        & [20, 20]$^{\dagger}$, not aligned & .8604 ± .0140 \\
        & [20, 20]$^{\dagger}$, aligned & .9007 ± .0052 \\
    \midrule

    \multirow{1}{*}{ [CSI0,CSI1]+CSI2} 
        & [20,80] + 80$^{\dagger}$ & \textbf{.9431} ± .0104  \\
    \bottomrule

\end{tabular}
}
\label{table:aligned_csi0_csi1}

{\raggedright
$^{\dagger}$ The notation \([{\rm BW}_0,{\rm BW}_1] + {\rm BW}_2\) indicates the fusion of the CSI clusters (CSI0, CSI1, CSI2) discovered in sanitization.\par

\par}

\end{table}

\noindent
{Fusion of CSI clusters within 5~GHz band}
{(Table~\ref{table:aligned_csi0_csi1}):}
Fusing heterogeneous clusters and bandwidths improves accuracy markedly. 
Aligning CSI0 and CSI1 (20\,MHz) boosts the performance. 
Adding CSI1 and CSI2 of wider bandwidth (80\,MHz) further lifts accuracy. 
This confirms that cross-cluster fusion plus waveform alignment are beneficial.

\begin{table}[t]
\centering
\caption{Accuracy of the CSI clusters in 2.4~GHz band.}
\resizebox{\columnwidth}{!}{%
\begin{tabular}{llcc}
    \toprule
    \textbf{CSI Cluster} & \textbf{Bandwidth (MHz)} & \textbf{Accuracy} \\ 
    \midrule
    
    \multirow{1}{*}{CSI0} 
        & 20 & .8937 ± .0092  \\
    \midrule

    \multirow{1}{*}{CSI1} 
        & 20 & .8857 ± .0351 \\
    \midrule

    \multirow{1}{*}{CSI2} 
        & 20 & .5357 ± .0322 \\
    \midrule

    \multirow{1}{*}{ [CSI0,CSI1]} 
        & [20, 20]$^{\dagger}$ & \textbf{.9056} ± .0102 \\
    \midrule

    \multirow{1}{*}{ [CSI0,CSI1]+CSI2} 
        & [20,20] + 20$^{\dagger}$ & .8611 ± .0128  \\
    \bottomrule

\end{tabular}
}
\label{table:node2_csi012}

{\raggedright
$^{\dagger}$ The notation \([{\rm BW}_0,{\rm BW}_1] + {\rm BW}_2\) indicates the fusion of the CSI clusters (CSI0, CSI1, CSI2) discovered in sanitization.\par

\par}
\end{table}

\noindent
{Fusion of CSI clusters within 2.4~GHz band}
{(Table~\ref{table:node2_csi012}):}
\rev{Table~\ref{table:node2_csi012} also provides a stress test of fusion robustness. CSI2 on 2.4\,GHz is the weakest standalone stream, which is consistent with the broader 2.4\,GHz versus 5\,GHz performance gap in Table~\ref{table:nodes_all}. Several factors contribute to this behavior. First, the 2.4\,GHz streams are limited to 20\,MHz bandwidth, providing fewer usable subcarriers and less frequency diversity than the strongest 5\,GHz streams. Second, CSI2 has weaker signal quality and class separability than CSI0/CSI1: its ACV is higher (0.298 versus 0.121/0.119), its 95th-percentile L2 distance from the clip mean is larger (0.438 versus 0.216/0.214), and its Fisher ratio is much lower (0.058 versus 0.403/0.394). Third, CSI2 is the QoS-data cluster and is sparse and bursty in our measurements (MR $=0.7909$, SCV $=2.0141$ in Table~\ref{table:csi_metrics}), placing it closer to the difficult sparse-and-irregular regime in Figure~\ref{fig:Heatmap_MR_SCV}.}

\rev{Despite the weak standalone CSI2 stream, fusing it with the cleaner 2.4\,GHz clusters degrades accuracy only marginally rather than catastrophically. Dual-band fusion is evaluated separately and further improves overall performance by exploiting complementary information across bands. This suggests that \ourSystem{} can tolerate a compromised or weak stream by relying more on informative observations from other clusters, supporting the robustness of the cluster-fusion design.}

\subsubsection{\textbf{\rev{Robustness to Hyperparameters}}}
\rev{We evaluated FuseFi-DNN's hyperparameter robustness on the dual-band CommCSI-HAR configuration by varying one hyperparameter at a time around the default ($h{=}64$, $d_e{=}64$, $Q{=}64$, $H{=}1$, $\eta{=}10^{-3}$). Across the sweep, final-epoch accuracy remains within 96.1--98.1\%. Varying the time-embedding dimension ($d_e{=}32$--128) and reference query points ($Q{=}32$--256) has minimal effect, while increasing attention heads or hidden size provides modest accuracy gains at the cost of additional computation or parameters. The default configuration achieves 96.88\%, within about 1.2 percentage points of the best variant, while using 3.7$\times$ fewer parameters than the largest hidden-size setting. This indicates that FuseFi does not rely on finely tuned hyperparameters and that the default configuration provides a favorable accuracy-efficiency trade-off.}

\subsection{Communication Cost}
\label{subsec:comm_cost}

We note that \ourSystem{} introduces \rev{zero sensing-specific communication overhead on the monitored channel: it does not inject sensing packets and therefore does not add airtime, reduce throughput, or introduce extra co-channel interference. This should be distinguished from hardware overhead: the current prototype uses Raspberry Pi sniffers as measurement instruments for CSI readout, whereas an integrated deployment would expose CSI through AP/client firmware or driver support.}
\ourSystem fuses all available CSI from diverse packet types, bandwidths, and frequency bands, even with irregular timestamps. Unlike prior works such as SenCom and CSI-BERT2, they may rely on active packet injection to maintain sensing fidelity. Such injections inevitably interfere with communication links in the same channel. For example, WiImg2~\cite{yang2022wiimg} has shown that active packet injection can reduce throughput by over 40\%.

\section{Discussion and Future Work}
\label{sec:discussion}

\revtwo{Overall, the current evaluation establishes FuseFi's effectiveness for irregular and heterogeneous communication CSI across multiple sensing tasks and dataset settings, while several deployment and generalization limitations remain. In particular, the subject-mixed CommCSI-HAR evaluation does not establish zero-shot cross-subject generalization, the evaluations across independently collected datasets demonstrate architectural generality rather than direct cross-environment transfer of a single trained model, and adaptive threshold tuning and a complete end-to-end edge implementation remain future work. These limitations and corresponding directions are discussed in Sections~VI-D--VI-F.}

\subsection{Integrated with Existing Hardware}
Similar to existing Wi-Fi sensing studies~\cite{he2023sencom,zhang2021widar3,zhao2024finding,zhao2024mining,portner2026same}, our current prototype uses separate COTS devices as sniffers to capture CSI. 
\rev{This is a prototyping choice for flexible CSI collection rather than an architectural requirement of FuseFi. FuseFi requires only per-packet CSI, or equivalent per-packet channel estimates, from naturally occurring communication traffic. In an integrated deployment, such channel estimates could be exposed directly by the AP or client through firmware or driver support, as demonstrated by prior tools such as Nexmon~\cite{schulz2018nexmon}. Since these estimates are already computed at the receiver PHY for normal equalization, exposing them for sensing does not add radio resources, airtime, or sensing-specific packets.}

\subsection{CSI Phase Utilization}
\ourSystem{} currently focuses on CSI amplitude and does not explicitly use phase information. This choice is deliberate. While CSI phase can contain additional channel information, raw phase from COTS Wi-Fi devices is strongly affected by carrier-frequency offset, sampling-frequency offset, packet-detection delay, and random per-packet phase offsets~\cite{xie2015precise,hu2024you}. 
\rev{These effects are especially difficult to calibrate when CSI is collected opportunistically from heterogeneous packet types, multiple bandwidths, multiple bands, and independent sniffers. CSI amplitude is less sensitive to such impairments and is therefore a robust default signal for the injection-free, asynchronous, multi-band communication-CSI setting studied in this paper. Amplitude-only CSI is also widely used in COTS Wi-Fi HAR tasks~\cite{hu2024you,zhao2024finding,zhao2024knn,zhao2024mining} such as activity recognition, gesture recognition, fall detection, and people counting.}

\rev{At the same time, phase, Doppler, AoA, and multi-antenna features can provide richer motion and spatial information when synchronized antennas or RF chains are available. Techniques such as reference-antenna conjugate multiplication, CSI-ratio processing, Doppler extraction, and AoA estimation are complementary to \ourSystem{}. Since \ourSystem{} operates on irregular timestamped CSI-derived observations, calibrated phase, Doppler, AoA, or multi-antenna features could be appended as additional channels and fused by the same time-aware architecture. We leave these extensions as future work.}

\subsection{Beyond Classification Tasks}
This paper primarily evaluates FuseFi on multiple classification-oriented sensing tasks, including HAR, gesture recognition, fall detection, and people counting. Other Wi-Fi sensing applications, such as localization and imaging, are also important but may impose different signal and modeling requirements. In particular, such tasks may benefit more from spatial or phase-related information, whereas FuseFi currently focuses on amplitude for robustness under heterogeneous communication CSI. Extending FuseFi to these applications is a promising direction for future work.

\subsection{Subject and Environmental Generalization}
\label{subsubsec:env_gen}
\rev{\textbf{Subject generalization.}
This paper focuses on communication-driven sensing from irregular and heterogeneous CSI, and does not establish subject-independent generalization for a single trained model. 
As shown in Section~\ref{sec:evaluation}, leave-one-subject-out evaluation reveals a gap between in-domain and cross-subject performance, with similar trends observed on WiFall-Action and with CSI-BERT2. This suggests that zero-shot cross-subject Wi-Fi-CSI HAR remains broadly challenging.
One reason is the limited number of subjects in existing datasets (8 in CommCSI-HAR and 10 in WiFall-Action), so each LOSO fold has only 7--9 training subjects. In addition, subject identity can be partially coupled with collection sessions, during which multipath, ambient interference, device placement, and channel occupancy may vary~\cite{cominelli2023exposing}. Thus, LOSO can also act as a more stringent out-of-session test.}

\rev{A small amount of user-specific calibration substantially improves performance: with 5--10 labeled windows per activity, FuseFi reaches approximately 91.3\%--94.7\% accuracy. This suggests that lightweight calibration is a practical deployment path, while calibration-free deployment via cross-domain adaptation~\cite{chen2023cross} and unsupervised/test-time adaptation~\cite{chen2023robust,chen2022fidora} remains an important future direction.}

\rev{\textbf{Environmental generalization.}
This paper does not evaluate cross-environment generalization across rooms, layouts, AP placements, or ambient interference conditions. Environment-independent Wi-Fi sensing is a complementary cross-domain problem, with dedicated techniques such as domain adaptation and environment-invariant representation learning~\cite{zhang2021widar3,meneghello2022sharp,gao2021towards, abuhoureyah2025location,liu2025genfi}. FuseFi addresses a different but compatible problem: extracting sensing information from naturally occurring, irregularly sampled communication CSI across heterogeneous packet types and frequency bands. Future work should validate FuseFi across multiple rooms, layouts, AP placements, and interference conditions, and can combine our communication-CSI preprocessing and time-aware fusion pipeline with environment-adaptation modules.}

\rev{\subsection{Adaptive Threshold Tuning}}
\rev{The sanitization thresholds $\tau_d$, $T_b$, and $T_c$ are system design parameters with different roles: $\tau_d$ controls CSI outlier pruning within PHY/bandwidth clusters, $T_b$ controls burst-packet removal, and $T_c$ controls the temporal window for cross-cluster alignment. Although the sensitivity analysis in Section~\ref{subsubsec:threshold_sensitivity} shows stable performance across practical ranges, these thresholds could be adapted online using traffic statistics. For example, $\tau_d$ could be set from short-window cluster-stability statistics, $T_b$ from the observed inter-packet-interval distribution, and $T_c$ from the availability of near-coincident packets across clusters. Since MR and SCV can also be estimated online, a deployment could monitor these statistics and adjust $T_b$ and $T_c$ to avoid high-missingness and high-irregularity regimes. We leave a full online adaptive-threshold implementation as future work.}

\rev{\subsection{Edge-Deployment Feasibility}}
\label{subsec:edge_deploy}
\revtwo{Model training is performed offline on the server described in Section~V-B. Here, we focus on evaluating the neural-network inference efficiency} of FuseFi and CSI-BERT2 on a Raspberry Pi 5 as a representative Pi-class edge platform, using PyTorch CPU inference with batch size 1, the same 52-subcarrier $\times$ 100-sample input setting as in Section~\ref{sec:evaluation}, and 1000 inference runs per model. FuseFi has 155.8K parameters and requires approximately 4.5\,MFLOPs per inference window, whereas CSI-BERT2 has 5.45M parameters and requires approximately 126\,MFLOPs. On the Raspberry Pi 5, FuseFi runs in 8.65$\pm$0.24\,ms per inference window, compared with 16.37$\pm$0.59\,ms for CSI-BERT2. FuseFi also uses substantially less runtime memory (1.28\,MB versus 22.1\,MB) and lower energy per inference (35.2$\pm$0.3\,mJ versus 80.8$\pm$2.0\,mJ). These measurements show that FuseFi's compactness translates into practical inference advantages on resource-constrained devices, beyond smaller theoretical FLOPs.

\revtwo{The reported Raspberry Pi 5 measurements characterize only the neural-network inference stage.} In the current prototype, CSI sanitization, which consists of conventional signal-processing operations commonly used in CSI sensing pipelines~\cite{schulz2018nexmon, portner2026same, gringoli2019free}, is implemented offline in MATLAB R2021b. 
\revtwo{A complete real-time edge implementation integrating CSI acquisition, sanitization, and inference is left for future work.}
A practical deployment could train FuseFi on a server, export the compact model weights, and execute neural inference on an edge device using preprocessed CSI sensing windows.

\subsection{Broader ISAC Context and Future Extensions}
This paper mainly focuses on ISAC studies that exploit existing communication packets for sensing. More broadly, however, ISAC spans a wider design space. For example, Baton~\cite{zhao2025baton} investigates device-free tracking by leveraging signals from multiple links. However, Baton still assumes fixed-rate sensing with injected packets at 1000~Hz, whereas our work focuses on classification tasks using irregularly sampled, heterogeneous CSI derived directly from communication packets. Extending FuseFi to multi-AP or multi-link settings is therefore a natural future direction that could further improve robustness and coverage through spatial diversity.
\rev{Wi-Fi~7 Multi-Link Operation (MLO) is used in this paper as a motivating industry trend toward multi-band connectivity, not as an implementation requirement of FuseFi. The present prototype uses two independent NTP-synchronized sniffers to capture 2.4\,GHz and 5\,GHz traffic and does not implement the MLO protocol. As discussed in Section~\ref{sec:evaluation}, FuseFi's amplitude-based cross-band fusion is robust to synthetic cross-band timing skew, so it does not require phase-coherent MLO-style synchronization. MLO-capable hardware could provide a more native multi-band CSI source in future deployments.}
Moreover, although FuseFi focuses on strictly passive sensing, our pipeline is also compatible with hybrid sensing strategies that inject sparse probes during long idle periods and fuse the resulting CSI with communication-derived CSI across varying frame types, bandwidths, and irregular timestamps.

\section{Conclusion}

This paper presents \ourSystem, a practical Wi-Fi sensing solution for ISAC scenarios that leverages irregularly sampled CSI from existing communication packets without packet injection, thereby maximizing the utility of available CSI.
\ourSystem sanitizes heterogeneous CSI from diverse packet types, bandwidths, and frequency bands, and employs a time-aware DNN model to process the resulting irregular data streams.
Furthermore, we collect \textit{CommCSI-HAR}, the first HAR dataset featuring irregularly sampled CSI extracted from real communication traffic across dual frequency bands using commodity Wi-Fi devices.
Extensive experiments across five representative sensing tasks validate its effectiveness and efficiency, \revtwo{while also highlighting the remaining challenges in zero-shot cross-subject generalization and cross-environment transfer.}
\revtwo{Addressing these generalization challenges together with fully integrated real-time edge deployment remains important future work.}

\section{Acknowledgment}
The research reported in this paper was sponsored in part by: the Air Force Office of Scientific Research under award \#FA95502310559; the DEVCOM Army Research \rev{Laboratory} under award \#W911NF1720196; the National Science Foundation under awards \#CNS-2211301 and ECCS-2525614; and, the National Institutes of Health under award \#1P41EB028242. The views and conclusions contained in this document are those of the authors and should not be interpreted as representing the official policies, either expressed or implied, of the funding agencies.

\bibliographystyle{IEEEtran}
\bibliography{references}

@article{huang2020iot,
  title={{IoT Inspector: Crowdsourcing Labeled Network Traffic from Smart Home Devices at Scale}},
  author={Huang, Danny Yuxing and Apthorpe, Noah and Li, Frank and Acar, Gunes and Feamster, Nick},
  journal={Proceedings of the ACM on Interactive, Mobile, Wearable and Ubiquitous Technologies},
  volume={4},
  number={2},
  pages={1--21},
  year={2020},
  publisher={ACM New York, NY, USA}
}

@inproceedings{mazhar2020characterizing,
  title={{Characterizing Smart Home IoT Traffic in the Wild}},
  author={Mazhar, M Hammad and Shafiq, Zubair},
  booktitle={IEEE/ACM IoTDI},
  xpages={203--215},
  year={2020},
  xorganization={IEEE}
}

@inproceedings{mainuddin2022iot,
  title={{IoT Device Identification Based on Network Traffic Characteristics}},
  author={Mainuddin, Md and Duan, Zhenhai and Dong, Yingfei and Salman, Shaeke and Taami, Tania},
  booktitle={IEEE GLOBECOM},
  xpages={6067--6072},
  year={2022},
  xorganization={IEEE}
}

@article{zheng2024pushing,
  title={Pushing the limits of WiFi sensing with low transmission rates},
  author={Zheng, Xiaolong and Yang, Kun and Xiong, Jie and Liu, Liang and Ma, Huadong},
  journal={IEEE Transactions on Mobile Computing},
  year={2024},
  publisher={IEEE}
}

@inproceedings{yang2022wiimg,
  title={Wiimg: Pushing the limit of wifi sensing with low transmission rates},
  author={Yang, Kun and Zheng, Xiaolong and Xiong, Jie and Liu, Liang and Ma, Huadong},
  booktitle={2022 19th Annual IEEE International Conference on Sensing, Communication, and Networking (SECON)},
  pages={1--9},
  year={2022},
  organization={IEEE}
}

@article{wang2023automatic,
  title={Automatic update for wi-fi fingerprinting indoor localization via multi-target domain adaptation},
  author={Wang, Jiankun and Zhao, Zenghua and Ou, Mengling and Cui, Jiayang and Wu, Bin},
  journal={Proceedings of the ACM on Interactive, Mobile, Wearable and Ubiquitous Technologies},
  volume={7},
  number={2},
  pages={1--27},
  year={2023},
  publisher={ACM New York, NY, USA}
}

@article{xu2023self,
  title={Self-supervised learning for wifi csi-based human activity recognition: A systematic study},
  author={Xu, Ke and Wang, Jiangtao and Zhu, Hongyuan and Zheng, Dingchang},
  journal={arXiv preprint arXiv:2308.02412},
  year={2023}
}

@inproceedings{cominelli2023exposing,
  title={Exposing the CSI: A systematic investigation of CSI-based Wi-Fi sensing capabilities and limitations},
  author={Cominelli, Marco and Gringoli, Francesco and Restuccia, Francesco},
  booktitle={2023 IEEE International Conference on Pervasive Computing and Communications (PerCom)},
  pages={81--90},
  year={2023},
  organization={IEEE}
}

@incollection{zhang2023wifi,
  title={Wifi/4g/5g based wireless sensing: Theories, applications and future directions},
  author={Zhang, Daqing and Niu, Kai and Xiong, Jie and Zhang, Fusang and Wang, Xuanzhi},
  booktitle={Integrated Sensing and Communications},
  pages={387--417},
  year={2023},
  publisher={Springer}
}

@article{liu2024unifi,
  title={Unifi: a unified framework for generalizable gesture recognition with wi-fi signals using consistency-guided multi-view networks},
  author={Liu, Yan and Yu, Anlan and Wang, Leye and Guo, Bin and Li, Yang and Yi, Enze and Zhang, Daqing},
  journal={Proceedings of the ACM on Interactive, Mobile, Wearable and Ubiquitous Technologies},
  volume={7},
  number={4},
  pages={1--29},
  year={2024},
  publisher={ACM New York, NY, USA}
}

@article{niu2021understanding,
  title={Understanding WiFi signal frequency features for position-independent gesture sensing},
  author={Niu, Kai and Zhang, Fusang and Wang, Xuanzhi and Lv, Qin and Luo, Haitong and Zhang, Daqing},
  journal={IEEE Transactions on Mobile Computing},
  volume={21},
  number={11},
  pages={4156--4171},
  year={2021},
  publisher={IEEE}
}

@article{deng2022gaitfi,
  title={GaitFi: Robust device-free human identification via WiFi and vision multimodal learning},
  author={Deng, Lang and Yang, Jianfei and Yuan, Shenghai and Zou, Han and Lu, Chris Xiaoxuan and Xie, Lihua},
  journal={IEEE Internet of Things Journal},
  volume={10},
  number={1},
  pages={625--636},
  year={2022},
  publisher={IEEE}
}

@article{ma2019wifi,
  title={WiFi sensing with channel state information: A survey},
  author={Ma, Yongsen and Zhou, Gang and Wang, Shuangquan},
  journal={ACM Computing Surveys (CSUR)},
  volume={52},
  number={3},
  pages={1--36},
  year={2019},
  publisher={ACM New York, NY, USA}
}

@book{gast2013802,
  title={802.11 ac: A survival guide},
  author={Gast, Matthew},
  year={2013},
  publisher={" O'Reilly Media, Inc."}
}

@article{shukla2021multi,
  title={Multi-time attention networks for irregularly sampled time series},
  author={Shukla, Satya Narayan and Marlin, Benjamin M},
  journal={arXiv preprint arXiv:2101.10318},
  year={2021}
}

@article{schulz2018nexmon,
  title={The Nexmon firmware analysis and modification framework: Empowering researchers to enhance Wi-Fi devices},
  author={Schulz, Matthias and Wegemer, Daniel and Hollick, Matthias},
  journal={Computer Communications},
  volume={129},
  pages={269--285},
  year={2018},
  publisher={Elsevier}
}

@article{yang2023sensefi,
  title={SenseFi: A library and benchmark on deep-learning-empowered WiFi human sensing},
  author={Yang, Jianfei and Chen, Xinyan and Zou, Han and Lu, Chris Xiaoxuan and Wang, Dazhuo and Sun, Sumei and Xie, Lihua},
  journal={Patterns},
  volume={4},
  number={3},
  year={2023},
  publisher={Elsevier}
}

@article{abedi2023wifi,
  title={WiFi physical layer stays awake and responds when it should not},
  author={Abedi, Ali and Lu, Haofan and Chen, Alex and Liu, Charlie and Abari, Omid},
  journal={IEEE Internet of Things Journal},
  volume={11},
  number={3},
  pages={4483--4496},
  year={2023},
  publisher={IEEE}
}

@article{zhang2019widetect,
  title={WiDetect: Robust motion detection with a statistical electromagnetic model},
  author={Zhang, Feng and Wu, Chenshu and Wang, Beibei and Lai, Hung-Quoc and Han, Yi and Liu, KJ Ray},
  journal={Proceedings of the ACM on Interactive, Mobile, Wearable and Ubiquitous Technologies},
  volume={3},
  number={3},
  pages={1--24},
  year={2019},
  publisher={ACM New York, NY, USA}
}

@article{zhang2019smars,
  title={SMARS: Sleep monitoring via ambient radio signals},
  author={Zhang, Feng and Wu, Chenshu and Wang, Beibei and Wu, Min and Bugos, Daniel and Zhang, Hangfang and Liu, KJ Ray},
  journal={IEEE Transactions on Mobile Computing},
  volume={20},
  number={1},
  pages={217--231},
  year={2019},
  publisher={IEEE}
}

@article{palipana2018falldefi,
  title={FallDeFi: Ubiquitous fall detection using commodity Wi-Fi devices},
  author={Palipana, Sameera and Rojas, David and Agrawal, Piyush and Pesch, Dirk},
  journal={Proceedings of the ACM on Interactive, Mobile, Wearable and Ubiquitous Technologies},
  volume={1},
  number={4},
  pages={1--25},
  year={2018},
  publisher={ACM New York, NY, USA}
}

@article{gao2021towards,
  title={Towards position-independent sensing for gesture recognition with Wi-Fi},
  author={Gao, Ruiyang and Zhang, Mi and Zhang, Jie and Li, Yang and Yi, Enze and Wu, Dan and Wang, Leye and Zhang, Daqing},
  journal={Proceedings of the ACM on Interactive, Mobile, Wearable and Ubiquitous Technologies},
  volume={5},
  number={2},
  pages={1--28},
  year={2021},
  publisher={ACM New York, NY, USA}
}

@article{zhao2024mining,
  title={{CSI-BERT2: A BERT-inspired Framework for Efficient CSI Prediction and Classification in Wireless Communication and Sensing}},
  author={Zhao, Zijian and Meng, Fanyi and Li, Hang and Li, Xiaoyang and Zhu, Guangxu},
  journal={arXiv preprint arXiv:2412.06861},
  year={2024}
}

@inproceedings{he2023sencom,
  title={Sencom: Integrated sensing and communication with practical wifi},
  author={He, Yinghui and Liu, Jianwei and Li, Mo and Yu, Guanding and Han, Jinsong and Ren, Kui},
  booktitle={Proceedings of the 29th Annual International Conference on Mobile Computing and Networking},
  pages={1--16},
  year={2023}
}

@article{liu2022integrated,
  title={Integrated sensing and communications: Toward dual-functional wireless networks for 6G and beyond},
  author={Liu, Fan and Cui, Yuanhao and Masouros, Christos and Xu, Jie and Han, Tony Xiao and Eldar, Yonina C and Buzzi, Stefano},
  journal={IEEE journal on selected areas in communications},
  volume={40},
  number={6},
  pages={1728--1767},
  year={2022},
  publisher={IEEE}
}

@article{zhang2024integrated,
  title={Integrated sensing and communication with massive MIMO: A unified tensor approach for channel and target parameter estimation},
  author={Zhang, Ruoyu and Cheng, Lei and Wang, Shuai and Lou, Yi and Gao, Yulong and Wu, Wen and Ng, Derrick Wing Kwan},
  journal={IEEE Transactions on Wireless Communications},
  volume={23},
  number={8},
  pages={8571--8587},
  year={2024},
  publisher={IEEE}
}

@article{li2025ris,
  title={RIS-based Physical Layer Security for Integrated Sensing and Communication: A Comprehensive Survey},
  author={Li, Yongxiao and Khan, Feroz and Ahmed, Manzoor and Soofi, Aized Amin and Khan, Wali Ullah and Sheemar, Chandan Kumar and Asif, Muhammad and Han, Zhu},
  journal={IEEE Internet of Things Journal},
  year={2025},
  publisher={IEEE}
}

@article{lu2024integrated,
  title={Integrated sensing and communications: Recent advances and ten open challenges},
  author={Lu, Shihang and Liu, Fan and Li, Yunxin and Zhang, Kecheng and Huang, Hongjia and Zou, Jiaqi and Li, Xinyu and Dong, Yuxiang and Dong, Fuwang and Zhu, Jia and others},
  journal={IEEE Internet of Things Journal},
  volume={11},
  number={11},
  pages={19094--19120},
  year={2024},
  publisher={IEEE}
}

@article{du2024overview,
  title={An overview on IEEE 802.11 bf: WLAN sensing},
  author={Du, Rui and Hua, Haocheng and Xie, Hailiang and Song, Xianxin and Lyu, Zhonghao and Hu, Mengshi and Xin, Yan and McCann, Stephen and Montemurro, Michael and Han, Tony Xiao and others},
  journal={IEEE Communications Surveys \& Tutorials},
  year={2024},
  publisher={IEEE}
}

@article{restuccia2021ieee,
  title={IEEE 802.11 bf: Toward ubiquitous Wi-Fi sensing},
  author={Restuccia, Francesco},
  journal={arXiv preprint arXiv:2103.14918},
  year={2021}
}

@article{chen2022wi,
  title={Wi-Fi sensing based on IEEE 802.11 bf},
  author={Chen, Cheng and Song, Hao and Li, Qinghua and Meneghello, Francesca and Restuccia, Francesco and Cordeiro, Carlos},
  journal={IEEE Communications Magazine},
  volume={61},
  number={1},
  pages={121--127},
  year={2022},
  publisher={IEEE}
}

@inproceedings{pegoraro2024hisac,
  title={HiSAC: high-resolution sensing with multiband communication signals},
  author={Pegoraro, Jacopo and Lacruz, Jesus O and Rossi, Michele and Widmer, Joerg},
  booktitle={Proceedings of the 22nd ACM Conference on Embedded Networked Sensor Systems},
  pages={549--563},
  year={2024}
}

@inproceedings{zhang2019commercial,
  title={Commercial Wi-Fi based fall detection with environment influence mitigation},
  author={Zhang, Lei and Wang, Zhirui and Yang, Liu},
  booktitle={2019 16th Annual IEEE International Conference on Sensing, Communication, and Networking (SECON)},
  pages={1--9},
  year={2019},
  organization={IEEE}
}

@inproceedings{shi2020towards,
  title={Towards environment-independent behavior-based user authentication using wifi},
  author={Shi, Cong and Liu, Jian and Borodinov, Nick and Leao, Bruno and Chen, Yingying},
  booktitle={2020 IEEE 17th International Conference on Mobile Ad Hoc and Sensor Systems (MASS)},
  pages={666--674},
  year={2020},
  organization={IEEE}
}

@inproceedings{li2016wifinger,
  title={WiFinger: Talk to your smart devices with finger-grained gesture},
  author={Li, Hong and Yang, Wei and Wang, Jianxin and Xu, Yang and Huang, Liusheng},
  booktitle={Proceedings of the 2016 ACM International Joint Conference on Pervasive and Ubiquitous Computing},
  pages={250--261},
  year={2016}
}

@article{guo2022twcc,
  title={TWCC: A robust through-the-wall crowd counting system using ambient WiFi signals},
  author={Guo, Zhengxin and Xiao, Fu and Sheng, Biyun and Sun, Lijuan and Yu, Shui},
  journal={IEEE Transactions on Vehicular Technology},
  volume={71},
  number={4},
  pages={4198--4211},
  year={2022},
  publisher={IEEE}
}

@inproceedings{zhao2024finding,
  title={Finding the missing data: A bert-inspired approach against package loss in wireless sensing},
  author={Zhao, Zijian and Chen, Tingwei and Meng, Fanyi and Li, Hang and Li, Xiaoyang and Zhu, Guangxu},
  booktitle={IEEE INFOCOM 2024-IEEE Conference on Computer Communications Workshops (INFOCOM WKSHPS)},
  pages={1--6},
  year={2024},
  organization={IEEE}
}

@article{zhao2024knn,
  title={KNN-MMD: Cross Domain Wireless Sensing via Local Distribution Alignment},
  author={Zhao, Zijian and Cai, Zhijie and Chen, Tingwei and Li, Xiaoyang and Li, Hang and Chen, Qimei and Zhu, Guangxu},
  journal={arXiv preprint arXiv:2412.04783},
  year={2024}
}

@inproceedings{xiong2013securearray,
  title={Securearray: Improving wifi security with fine-grained physical-layer information},
  author={Xiong, Jie and Jamieson, Kyle},
  booktitle={Proceedings of the 19th annual international conference on Mobile computing \& networking},
  pages={441--452},
  year={2013}
}

@misc{wikipedia_beacon_frame,
  author       = "{Wikipedia contributors}",
  title        = "{Beacon frame -- Wikipedia}{}",
  year         = "2025",
  howpublished = "\url{https://en.wikipedia.org/wiki/Beacon_frame}",
}

@article{che2018recurrent,
  title={Recurrent neural networks for multivariate time series with missing values},
  author={Che, Zhengping and Purushotham, Sanjay and Cho, Kyunghyun and Sontag, David and Liu, Yan},
  journal={Scientific reports},
  volume={8},
  number={1},
  pages={6085},
  year={2018},
  publisher={Nature Publishing Group UK London}
}

@inproceedings{zheng2016smokey,
  title={Smokey: Ubiquitous smoking detection with commercial WiFi infrastructures},
  author={Zheng, Xiaolong and Wang, Jiliang and Shangguan, Longfei and Zhou, Zimu and Liu, Yunhao},
  booktitle={IEEE INFOCOM 2016-The 35th Annual IEEE International Conference on Computer Communications},
  pages={1--9},
  year={2016},
  organization={IEEE}
}

@inproceedings{xie2015precise,
  title={Precise power delay profiling with commodity WiFi},
  author={Xie, Yaxiong and Li, Zhenjiang and Li, Mo},
  booktitle={Proceedings of the 21st Annual international conference on Mobile Computing and Networking},
  pages={53--64},
  year={2015}
}

@inproceedings{hu2024you,
  title={What you need is a good CSI},
  author={Hu, Yuqian and Zhu, Guozhen and Wang, Wei-Hsiang and Wang, Beibei and Liu, KJ Ray},
  booktitle={Proceedings of the 30th Annual International Conference on Mobile Computing and Networking},
  pages={1608--1610},
  year={2024}
}

@article{hochreiter1997long,
  title={Long short-term memory},
  author={Hochreiter, Sepp and Schmidhuber, J{\"u}rgen},
  journal={Neural computation},
  volume={9},
  number={8},
  pages={1735--1780},
  year={1997},
  publisher={MIT press}
}

@misc{parks_smart_home,
  author       = {{Parks Associates}},
  title        = {Parks Associates to Share Smart Home Device Research at {Wi-Fi NOW}},
  howpublished = {\url{https://www.parksassociates.com/blogs/in-the-news/parks-associates-to-share-smart-home-device-research-at-wifi-now}},
  year         = {2025}, 
}

@inproceedings{ji2023construct,
  title={Construct 3d hand skeleton with commercial wifi},
  author={Ji, Sijie and Zhang, Xuanye and Zheng, Yuanqing and Li, Mo},
  booktitle={Proceedings of the 21st ACM Conference on Embedded Networked Sensor Systems},
  pages={322--334},
  year={2023}
}

@article{zhang2022handgest,
  title={HandGest: Hierarchical sensing for robust-in-the-air handwriting recognition with commodity WiFi devices},
  author={Zhang, Jie and Li, Yang and Xiong, Haoyi and Dou, Dejing and Miao, Chunyan and Zhang, Daqing},
  journal={IEEE Internet of Things Journal},
  volume={9},
  number={19},
  pages={19529--19544},
  year={2022},
  publisher={IEEE}
}

@inproceedings{xiong2015tonetrack,
  title={Tonetrack: Leveraging frequency-agile radios for time-based indoor wireless localization},
  author={Xiong, Jie and Sundaresan, Karthikeyan and Jamieson, Kyle},
  booktitle={Proceedings of the 21st Annual International Conference on Mobile Computing and Networking},
  pages={537--549},
  year={2015}
}

@article{chen2019m,
  title={M3: Multipath Assisted Wi-Fi Localization with a Single Access Point},
  author={Chen, Zhe and Zhu, Guorong and Wang, Sulei and Xu, Yuedong and Xiong, Jie and Zhao, Jin and Luo, Jun and Wang, Xin},
  journal={IEEE Transactions on Mobile Computing},
  volume={20},
  number={2},
  pages={588--602},
  year={2019},
  publisher={IEEE}
}

@inproceedings{li2025muceiver,
  title={$\mu$Ceiver-Fi: Exploiting Spectrum Resources of Multi-Link Receiver for Fine-Granularity Wi-Fi Sensing},
  author={Li, Xin and He, Yinghui and Luo, Jun},
  booktitle={Proceedings of the 31st Annual International Conference on Mobile Computing and Networking},
  pages={1045--1059},
  year={2025}
}

@article{yu2025lowdetrack,
  title={LowDetrack: A Human Detection and Tracking System for Wi-Fi Low Packet Rates},
  author={Yu, Aiwen and Yuan, Cheng and Gao, Chenwen and Tong, Xinyu and Liu, Xiulong and Xie, Xin and Chen, Jiancheng and Li, Keqiu},
  journal={IEEE Internet of Things Journal},
  year={2025},
  publisher={IEEE}
}

@article{zeng2020multisense,
  title={MultiSense: Enabling multi-person respiration sensing with commodity WiFi},
  author={Zeng, Youwei and Wu, Dan and Xiong, Jie and Liu, Jinyi and Liu, Zhaopeng and Zhang, Daqing},
  journal={Proceedings of the ACM on Interactive, Mobile, Wearable and Ubiquitous Technologies},
  volume={4},
  number={3},
  pages={1--29},
  year={2020},
  publisher={ACM New York, NY, USA}
}

@article{yousefi2017survey,
  title={A survey on behavior recognition using WiFi channel state information},
  author={Yousefi, Siamak and Narui, Hirokazu and Dayal, Sankalp and Ermon, Stefano and Valaee, Shahrokh},
  journal={IEEE Communications Magazine},
  volume={55},
  number={10},
  pages={98--104},
  year={2017},
  publisher={IEEE}
}

@article{zhang2021widar3,
  title={Widar3. 0: Zero-effort cross-domain gesture recognition with Wi-Fi},
  author={Zhang, Yi and Zheng, Yue and Qian, Kun and Zhang, Guidong and Liu, Yunhao and Wu, Chenshu and Yang, Zheng},
  journal={IEEE Transactions on Pattern Analysis and Machine Intelligence},
  volume={44},
  number={11},
  pages={8671--8688},
  year={2021},
  publisher={IEEE}
}

@article{meneghello2022sharp,
  title={SHARP: Environment and person independent activity recognition with commodity IEEE 802.11 access points},
  author={Meneghello, Francesca and Garlisi, Domenico and Dal Fabbro, Nicol{\`o} and Tinnirello, Ilenia and Rossi, Michele},
  journal={IEEE Transactions on Mobile Computing},
  volume={22},
  number={10},
  pages={6160--6175},
  year={2022},
  publisher={IEEE}
}

@article{wang2024xrf55,
  title={Xrf55: A radio frequency dataset for human indoor action analysis},
  author={Wang, Fei and Lv, Yizhe and Zhu, Mengdie and Ding, Han and Han, Jinsong},
  journal={Proceedings of the ACM on Interactive, Mobile, Wearable and Ubiquitous Technologies},
  volume={8},
  number={1},
  pages={1--34},
  year={2024},
  publisher={ACM New York, NY, USA}
}

@inproceedings{yuanoctonet,
  title={OctoNet: A Large-Scale Multi-Modal Dataset for Human Activity Understanding Grounded in Motion-Captured 3D Pose Labels},
  author={Yuan, Dongsheng and Zhang, Xie and Hou, Weiying and Lyu, Sheng and Yu, Yuemin and Yu, Luca Jiang-Tao and Li, Chengxiao and Wu, Chenshu},
  booktitle={The Thirty-ninth Annual Conference on Neural Information Processing Systems Datasets and Benchmarks Track},
  year={2025}
}

@article{yang2023mm,
  title={Mm-fi: Multi-modal non-intrusive 4d human dataset for versatile wireless sensing},
  author={Yang, Jianfei and Huang, He and Zhou, Yunjiao and Chen, Xinyan and Xu, Yuecong and Yuan, Shenghai and Zou, Han and Lu, Chris Xiaoxuan and Xie, Lihua},
  journal={Advances in Neural Information Processing Systems},
  volume={36},
  pages={18756--18768},
  year={2023}
}

@article{zhu2025csi,
  title={CSI-Bench: A Large-Scale In-the-Wild Dataset for Multi-task WiFi Sensing},
  author={Zhu, Guozhen and Hu, Yuqian and Gao, Weihang and Wang, Wei-Hsiang and Wang, Beibei and Liu, KJ},
  journal={arXiv preprint arXiv:2505.21866},
  year={2025}
}

@article{zhao2025baton,
  title={Baton: Compensate for Missing Wi-Fi Features for Practical Device-free Tracking},
  author={Zhao, Yiming and Meng, Xuanqi and Tong, Xinyu and Liu, Xiulong and Xie, Xin and Qu, Wenyu},
  journal={IEEE Transactions on Mobile Computing},
  year={2025},
  publisher={IEEE}
}

@article{portner2026same,
  title={Same Signal, Different Story: Demystifying Receiver Effects in Wi-Fi Channel State Information},
  author={Portner, Fabian and Gringoli, Francesco and Hollick, Matthias and Asadi, Arash},
  journal={IEEE Internet of Things Journal},
  year={2026},
  publisher={IEEE}
}

@article{abuhoureyah2025location,
  title={Location independent human activity recognition using self-training CSI-based techniques for wireless sensor networks},
  author={Abuhoureyah, Fahd Saad and Wong, Yan Chiew},
  journal={IEEE internet of Things Journal},
  year={2025},
  publisher={IEEE}
}

@inproceedings{liu2025genfi,
  title={GenFi: Enhancing WiFi-Based Human Activity Recognition to Unseen Scenario Via Feature Disentanglement and Meta-Learning},
  author={Liu, Han and Li, Yi and Sikdar, Biplab},
  booktitle={ICC 2025-IEEE International Conference on Communications},
  pages={728--734},
  year={2025},
  organization={IEEE}
}

@inproceedings{brunzema2022gru,
  title={GRU-ODE-Bayes: Continuous modeling of sporadically-observed time series},
  author={De Brouwer, Edward and Simm, Jaak and Arany, Adam and Moreau, Yves},
  booktitle={Advances in Neural Information Processing Systems (NeurIPS)},
  year={2019}
}

@inproceedings{chen2023robust,
  title={Robust finger interactions with cots smartwatches via unsupervised siamese adaptation},
  author={Chen, Wenqiang and Wang, Ziqi and Quan, Pengrui and Peng, Zhencan and Lin, Shupei and Srivastava, Mani and Matusik, Wojciech and Stankovic, John},
  booktitle={Proceedings of the 36th Annual ACM Symposium on User Interface Software and Technology},
  pages={1--14},
  year={2023}
}

@article{chen2022fidora,
  title={Fidora: Robust WiFi-based indoor localization via unsupervised domain adaptation},
  author={Chen, Xi and Li, Hang and Zhou, Chenyi and Liu, Xue and Wu, Di and Dudek, Gregory},
  journal={IEEE Internet of Things Journal},
  volume={9},
  number={12},
  pages={9872--9888},
  year={2022},
  publisher={IEEE}
}

@article{chen2023cross,
  title={Cross-domain WiFi sensing with channel state information: A survey},
  author={Chen, Chen and Zhou, Gang and Lin, Youfang},
  journal={ACM Computing Surveys},
  volume={55},
  number={11},
  pages={1--37},
  year={2023},
  publisher={ACM New York, NY}
}

@article{gu2022wigrunt,
  title={WiGRUNT: WiFi-enabled gesture recognition using dual-attention network},
  author={Gu, Yu and Zhang, Xiang and Wang, Yantong and Wang, Meng and Yan, Huan and Ji, Yusheng and Liu, Zhi and Li, Jianhua and Dong, Mianxiong},
  journal={IEEE transactions on human-machine systems},
  volume={52},
  number={4},
  pages={736--746},
  year={2022},
  publisher={IEEE}
}

@inproceedings{gringoli2019free,
  title={Free your CSI: A channel state information extraction platform for modern Wi-Fi chipsets},
  author={Gringoli, Francesco and Schulz, Matthias and Link, Jakob and Hollick, Matthias},
  booktitle={Proceedings of the 13th International Workshop on Wireless Network Testbeds, Experimental Evaluation \& Characterization},
  pages={21--28},
  year={2019}
}

\begin{IEEEbiography}[{\includegraphics[
width=1in,
height=1.25in,
clip,
keepaspectratio
]
{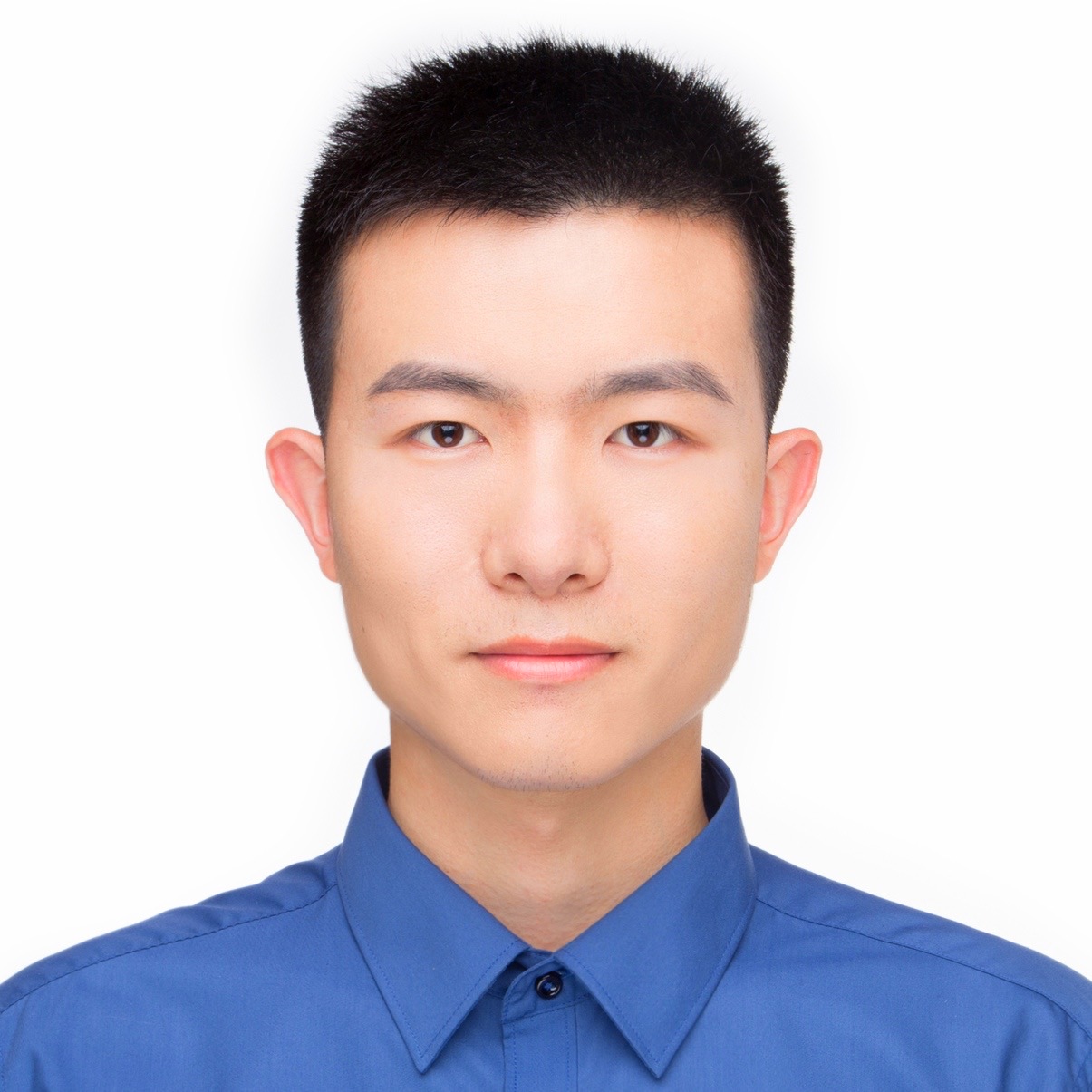}}]{Gaofeng Dong}
is currently a Ph.D. candidate with the Networked and Embedded Systems Laboratory (NESL) at the University of California, Los Angeles (UCLA), Los Angeles, CA, USA, advised by Prof. Mani B. Srivastava. He received the master’s and bachelor’s degrees in electrical engineering from the University of Science and Technology of China (USTC), Hefei, China, in 2020 and 2017, respectively. During his Ph.D. study, he interned at Bosch Research America in 2024. His research focuses on improving the efficiency, robustness, security, and privacy of embedded and IoT systems, with particular interests in mobile sensing and computing, deep learning, and foundation models for human–cyber–physical systems. His work has been published in leading venues including MobiSys, SenSys, NeurIPS, and IMWUT. He received the Best Paper Award for Foundation Models for Cyber–Physical Systems and Internet of Things at FMSys 2025.
\end{IEEEbiography}

\begin{IEEEbiography}[{\includegraphics[
width=1in,
height=1.25in,
clip,
keepaspectratio
]
{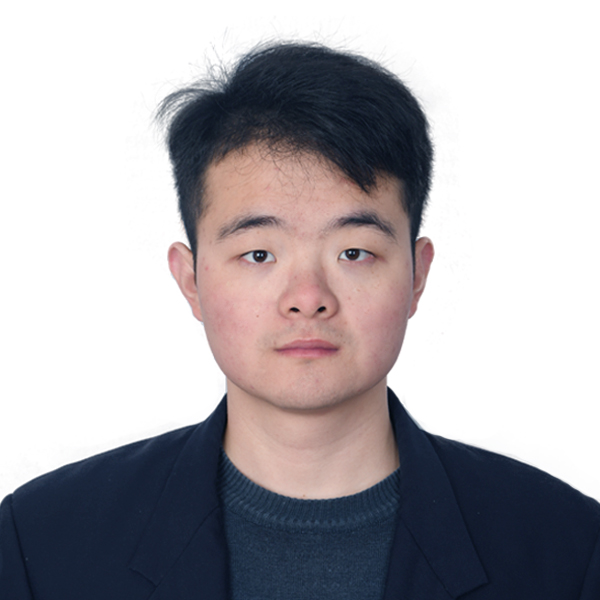}}]{Kang Yang}
is currently a Postdoctoral Scholar in the Department of Electrical and Computer Engineering, University of California, Los Angeles, CA, USA. He received the Ph.D. degree in electrical engineering and computer science from the University of California, Merced, CA, USA, in 2024. His research interests include AI-enabled design and optimization of cyber-physical systems.
\end{IEEEbiography}

\begin{IEEEbiography}[{\includegraphics[
width=1in,
height=1.25in,
clip,
keepaspectratio
]
{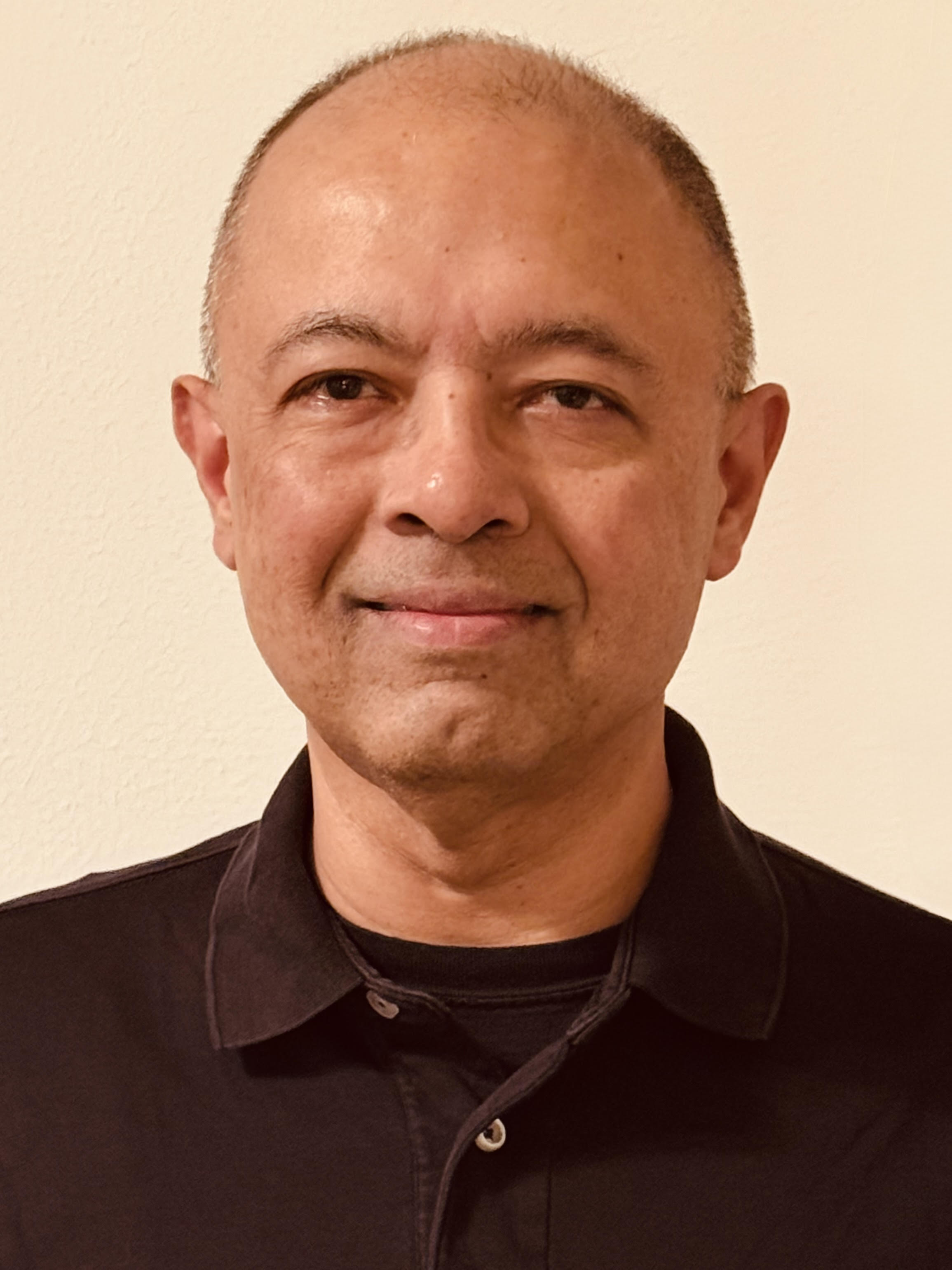}}]{Mani Srivastava}
Mani Srivastava is a Distinguished Professor, Mukund Padmanabhan Term Chair, and Vice Chair of Computer Engineering in the Department of Electrical and Computer Engineering at UCLA, with a joint appointment in Computer Science. His research focuses on learning-enabled, resource-efficient, and trustworthy human–cyber–physical and IoT systems, spanning applications in mobile health, smart environments, national security, and sustainability. He takes a full-stack approach, addressing challenges across the edge–cloud continuum, including application design, architectural abstractions, algorithms, and platform technologies. He is a Fellow of the ACM and IEEE.
\end{IEEEbiography}

\end{document}